\documentclass[11pt]{article}

\usepackage{multirow}
\usepackage{calc}
\usepackage{amsmath,amssymb,amsthm,amscd}
\numberwithin{equation}{section}
\usepackage[bf]{caption}
\usepackage{longtable}
\usepackage{array}
\usepackage{enumerate}
\usepackage{float}
\usepackage{subfig}
\usepackage{multicol}
\usepackage{epsfig}
\usepackage{graphicx}
\usepackage{color}
\usepackage{cite}
\usepackage[width=1.09\textwidth]{caption}
\usepackage{tikz}
\usepackage{hyperref}
\usepackage{tikz,textcomp}
\usetikzlibrary{shapes,arrows,positioning,chains,calc}

\usepackage[utf8]{inputenc}
\usepackage[T1]{fontenc}
\usepackage[english]{babel}
\usepackage{xspace}
\usepackage{pifont}

\newcommand{\beq}{\begin{equation}}
\newcommand{\eeq}{\end{equation}}
\newcommand{\bea}{\begin{eqnarray}}
\newcommand{\eea}{\end{eqnarray}}

\begin{document}

\baselineskip=15pt
\begin{titlepage}
 {\flushright 
 UUITP--11/21\\[5mm]} 

\begin{center}
\vspace*{ 2.0cm}
{\Large {\bf Non-flat elliptic four-folds, three-form cohomology and strongly coupled theories in four dimensions}}\\[12pt]
\vspace{-0.1cm}
\bigskip
\bigskip 
{
  {{Paul-Konstantin~Oehlmann} }
\bigskip }\\[3pt]
\vspace{0.cm}
{\it  
  Department of Physics and Astronomy,~Uppsala University,~Regementsv\"agen 1, 57120 Uppsala,~Sweden    
}
\\[2.0cm]
\end{center}

\begin{abstract}
\noindent In this note we consider smooth elliptic Calabi-Yau four-folds whose fiber ceases to be flat over compact Riemann surfaces of genus $g$ in the base. These non-flat fibers contribute K{\"a}hler moduli to the four-fold but also add to the three-form cohomology for $g>0$. In F-/M-theory these sectors are to be interpreted as compactifications of six/five dimensional $\mathcal{N}=(1,0)$ superconformal matter theories. 
The three-form cohomology leads to additional chiral singlets proportional to the dimension of five dimensional Coulomb branch of those sectors. We construct explicit examples for E-string theories as well as higher rank cases.
For the E-string theories
  we further investigate conifold transitions that remove those non-flat fibers.  
First, we show how non-flat fibers can be deformed from curves down to isolated points in the base. This removes the chiral singlet of the three-forms and leads to non-perturbative four-point couplings among matter fields which can be understood as remnants of the former E-string.  
Alternatively, the non-flat fibers can be avoided by performing birational base changes, analogous to 6D tensor branches. For compact bases these transitions alternate all Hodge numbers but leave the Euler number invariant.
\end{abstract}

\end{titlepage}
\clearpage
\setcounter{footnote}{0}
\setcounter{tocdepth}{2}
\tableofcontents
\clearpage

\section{Introduction}
In recent times, our understanding of strongly coupled super symmetric theories (possibly coupled to gravity) in various dimensions has made  great progress by means of geometric methods. Those theories are generically hard to control within the framework of regular quantum field theories. Six dimensional superconformal field theories (SCFT) for example are genuine strongly coupled and were believed to contain  massless string excitations in the spectrum \cite{Ganor:1996mu,Witten:1995gx,Witten:1995ex,Seiberg:1996vs,Bershadsky:1996nu}. Evidence for their existence could be made in explicit construction made in string theory. The string construction uses of compactification geometries that provide certain divisors that are wrapped by branes and lead to massive BPS strings in the non-compact directions.  If those divisors are collapsed to points these strings become tensionless and support the SCFT \cite{Witten:1996qb}. For 6D theories with minimal supersymmetry the most flexible tool of choice has been F-theory \cite{Vafa:1996xn}. F-theory itself geometrizes the type IIB axio-dilation into the complex structure of an auxiliary elliptic fiber which is put on top of the physical compactification space. The total geometry becomes that of an elliptic Calabi-Yau(CY) three-fold which features
the consistent description e.g. of branes with large IIB string coupling. Classifying all non-compact elliptic-threefolds that can be shrunk to a (possibly singular) point has lead to an extensive list of 6D SCFTs \cite{ Heckman:2013pva,DelZotto:2014hpa,Heckman:2015bfa} (see \cite{Heckman:2018jxk} for a review). This direction is also fruitful to further classify 5D SCFTs using M-theory on the same class of geometries, related to F-theory by a circle compactification \cite{Jefferson:2018irk,Bhardwaj:2019fzv}. To access M-theory though the CY geometry must be fully resolved. This can be done by performing birational base changes until the elliptic fiber admits at most minimal singularities.
Such fiber singularities then admit a crepant resolution according to the Tate-algorithm. This resolution strategy naturally reflects the 6D tensor branch  followed by the 5D coulomb branch upon the circle reduction.  However, there is yet another way to resolve the geometry. This resolution is crepant and hence Calabi-Yau but does not respect the dimension of the elliptic fiber, called non-flat resolution. In such cases a non-minimal fiber singularity in codimension two can be replaced by surfaces $E_i$ of complex dimension two.  Shrinking those surfaces back to points reaches the respective 5D SCFT point. This point of view has proven to be very efficient to  characterize 5D SCFTs and its phase structure \cite{Apruzzi:2018nre,Apruzzi:2019vpe,Apruzzi:2019opn,Hubner:2020uvb}. In addition it also allows to study the non-minimal singularities of elliptic three-folds and their F-theory lifts directly, without the need to change the base
  \cite{Buchmuller:2017wpe,Dierigl:2018nlv,Paul-KonstantinOehlmann:2019jgr}.  
Extending this geometric approach to 4D is of course highly desirable. Related geometric approaches e.g. those pioneered by Seiberg and Witten \cite{Seiberg:1994rs,Seiberg:1994aj} have lead to many new insights in $\mathcal{N}=2,3,4$ SCFTs and their construction \cite{Argyres:2015ffa, Caorsi:2018zsq,Garcia-Etxebarria:2016erx,Apruzzi:2020pmv}. Clearly, one would like to extend this program further to minimal supersymmetric theories in 4D. The reduced amount of supersymmetry though, does not protect the moduli spaces from non-perturbative corrections anymore which can obstruct the SCFT point. \\
This was also observed e.g. in \cite{Apruzzi:2018oge} where it was the goal to construct non-trivial SCFTs 
using F-theory on non-compact  elliptic four-folds similar extending the approach of superconformal matter \cite{DelZotto:2014hpa}.
Using birational base changes loci of non-minimal singularities in codimension three could be removed which introduced new K{\"a}hler moduli. However, the existence of this SCFT point at the origin of the K{\"a}hler
 moduli space might be obstructed by the aforementioned quantum corrections. The corrections originate from Euclidean D3 instantons that wrap the collapsing cycle and mix  K{\"a}hler and complex structure moduli space of the geometry.
 
This note wants to take a similar approach to \cite{Apruzzi:2018oge} and investigate the (classical) deformation space of elliptic four-folds with non-minimal singularities but in codimension two. Opposed to \cite{Apruzzi:2018oge} we do not want to focus solely on the resolution phase via a birational base change but also on the non-flat resolution and  (partial) Higgs branches.  A major difference of elliptic four-folds as opposed to three-folds is that they can have non-trivial three-form cohomology. As we will show in the following, these cohomology classes are naturally related to non-flat fibers in codimension two. The dimension of this moduli space is given by the additional independent Hodge number $h^{2,1}$. Due to their absence in three-folds and the fact that their dimension is self-dual under mirror symmetry \cite{Klemm:1996ts} makes the three-fold cohomology interesting in their own right. Unfortunately, these contributions have not been given much attention in the literature apart from\cite{Greiner:2015mdm,Greiner:2017ery}. Moreover, in the application to 4D SCFTs, this contribution yields yet another part in the moduli space of the four-fold which might mix with K{\"a}hler and complex structure moduli under quantum corrections.   

The non-flat contributions we are considering in this work appear in codimension two which makes them very analogous to the superconformal matter in three-folds. Indeed, one might construct four-folds simply from a three-fold fibered over another $\mathbb{P}^1$ which is a perspective we will large use in this work. Due to this analogy one might wonder weather it is possible to characterize the geometric moduli space of those non-flat fibrations via the 6D/5D SCFT data.   \\
Moreover, we want to show that these fibers and hence their non-perturbative contributions can dynamically be created/destroyed by studying conifold transition between four-folds.  
These transitions generically change the $h^{2,1}$ contribution but in fact also all other Hodge numbers. However we will find that transition related to base changes keep the Euler number invariant, which is important in order to connecting 4D F-theory vacua \cite{Intriligator:2012ue,Jockers:2016bwi}. \\
Another class of conifold transitions can be interpreted as a (partial) Higgs branch that keeps the gauge group but removes  
the non-flat fiber from codimension two. This transition does not fully avoid the non-flat fiber though. Instead they are pushed down to codimension three and hence correspond to points in the base of the four-fold.  Scenarios like those were considered in  \cite{Achmed-Zade:2018idx} where they were shown to lead to non-perturbative four-point couplings. These superpotential terms are mediated by D1 string instantons between matter curves that meet at the respective points. Those couplings are also present in our cases and involve matter curves that are remnants of the original E-string curve. Moreover, since the same transition admits an origin from a 6D partial Higgs branch  the involved matter curve representations of the four-point coupling are enforced geometrically and dictated by 6D anomaly cancellation. 

This work is structured as follows: In Section~\ref{sec:two} we give an overview of the geometry of elliptic four-folds. We infer the Hodge numbers for non-flat fibrations and link those in Section~\ref{ssec:4612} to the 6D/5D SCFT data. Those contribute additional chiral singlets in a four-fold which we review in Section~\ref{ssec:HodgeNumbers}. 
In Section~\ref{ssec:HiggsCouplings} we also show how (partial) Higgs branches of those SCFTs naturally to the prediction of non-perturbative four-point couplings. In Section~\ref{sec:three} we present our two main examples which are families of four-folds with E-string theories that exhibit a $(E_6 \times U(1))/\mathbb{Z}_3$ and $(SO(10) \times U(1)^2)/\mathbb{Z}_4$ gauge group. Our main tool to study and construct these compact CY geometries is toric geometry. However we also expect our arguments to be valid beyond those constructions and for non-compact geometries as we. In Section~\ref{sec:four} we also give examples with higher rank SCFTs that have an $E_8 \times SU(3) \times SU(2)$ gauge group, to show the validity of our proposal. More details of the toric resolutions are presented in the Appendices~\ref{app:E6toric}.  
\section{Geometric Preliminaries and F-theory}
\label{sec:two} 
One of the main goals of this note is to give a physical explanation to certain Hodge numbers of elliptic Calabi-Yau four-fold $X_4$ with non-flat fibers in the context of F/M-theory. For this we start by reviewing the independent Hodge numbers of a Calabi-Yau four-fold. These are depicted in the upper half of the Hodge diamond in Figure~\ref{fig:HodgeDiamondX4}.
A four-fold $X_4$ admits  K{\"a}hler and complex structure parameters whose dimensions are counted by $h^{1,1}(X_4)$ and $h^{3,1}(X_4)$ analogous to those found in three-folds.
 Unlike three-folds, four-folds can also exhibit non-trivial three-form cohomology counted by $h^{2,1}(X_4)$ which is self-dual under mirror symmetry.
\begin{figure}[t!]
\begin{center}
\begin{tabular}{ccccccccc}
&&&&1&&&& \\
&&&0&&0&&& \\
&&0&&$h^{1,1}$&&0&& \\
  &0&&$h^{2,1}$&&$h^{2,1}$&&0& \\
1  &&$h^{3,1}$&&$h^{2,2}$&&$h^{3,1}$&&1 \\
\end{tabular}
\caption{{\it \label{fig:HodgeDiamondX4}The Hodge half-diamond of a CY four-fold and its non-trivial entries. $h^{2,2}$ is not independent but related to the other three entries.}}
\end{center}
\end{figure}
The dimension of $h^{2,2}(X_4)$ of a four-fold is not  independent but related to the other Hodge numbers  as
\begin{align}
h^{2,2} = 2 (22+2 h^{1,1}+2 h^{3,1}-h^{2,1}) \, .
\end{align}
The Euler number can be computed as
\begin{align}
\chi  = 6 (8+  h^{1,1} +   h^{3,1} -   h^{2,1}) \, .
\end{align}
This quantity will be important later when discussing transitions among smooth four-folds.  
In addition we are interested in elliptic four-folds with non-flat fibers in codimension two.

In order to investigate the contribution of non-flat fibers to the K{\"a}hler moduli we first turn to smooth three-folds $X_3$. 
For a smooth elliptic three-fold $X_3$ we are able to split up the K{\"a}hler moduli contributions into pieces that directly connect with the F-theory interpretation. This split is  due to the Shioda-Tate-Wazir theorem and given as
\begin{align}
\label{eq:stwX3}
h^{1,1}(X_3)= 1+ \text{rank}(MW(X_3))+n_{\text{fibral}}+h^{1,1}(B_2)   \, .
\end{align}
The first contribution is that of the Mordell-Weil group and the number of $n_{\text{fibral}}$ fibral divisors 
 that resolve some ABCDEFG type singularity. Taken together, both parts count the rank of the full F-theory gauge group.
 Second there is the contribution from base divisors that contribute the  universal hypermultiplet and the 6D tensor multiplets.
   In \cite{Buchmuller:2017wpe} it was first proposed that non-flat fibrations generally require the addition of new divisor classes in that decomposition. In practice there can be $k=1\ldots n_{p}$ points in the two-fold base $B_2$ where the fiber decomposes into $E_{i,k}$ surface components with $i=1 \ldots n_{\text{non-flat},k}$ \cite{Buchmuller:2017wpe,Dierigl:2018nlv,Apruzzi:2019opn} as we will further explain in Section~\ref{ssec:4612}.
   
Having reviewed the three-fold case, we now turn to four-folds.
 Inspired from eqn.~\ref{eq:stwX3} we propose a similar type of Shioda-Tate-Wazir decomposition for four-folds given by
\begin{align}
\label{eq:hodgesplit}
h^{1,1}(X_4) =1+\text{rank}(MW(X_4)) +  h^{1,1}(B_3)+  n_{\text{fibral}}   + h^{1,1}(X_4)_{\text{non-flat}}\, .
\end{align}
The general decomposition is analogous to the three-fold case but we explicitly added contributions from non-flat fibers.
In the case of four-folds these are $n_{c}$ codimension two loci in $B_3$ and as such irreducible Riemann surfaces $\mathcal{C}_{\alpha}$ of genus $g_\alpha$ with $\alpha =1 \ldots  n_{c}$. Just as for three-folds, these curves admit complex two dimensional surfaces  $E_{i,\alpha}$ as their fiber. In analogy to three-folds these non-flat fibers $E_{i,\alpha}$  can be shrunk to singular curves $\mathcal{C}_\alpha$ in the four-fold geometry controlled by a volume modulus of the surface. Hence these surfaces contribute a additional K{\"a}hler parameters to those of the four-fold. Note that there can be additional  non-flat degenerations in codimension three on the base $B_3$. These types of non-flat fibers are rather generic and do not need to contribute to $h^{1,1}(X_4)$ in the decomposition of Equation~\ref{eq:hodgesplit}. The degenerations in codimension three that we are going to study here e.g. in Sections~\ref{sec:three} and throughout this paper are relatively mild and indeed do not contribute to $h^{1,1}(X_4)$.  

In the following we want to deduce how non-flat fibers contribute to $h^{2,1}(X_4)$. Non-flat fibers will in general not be the only source to the three-form cohomology but the one we want to focus on in this work. As we have just extended the non-flat surfaces $E_{i,\alpha}$ to be fibered over a curve their total space is a divisor divisors $D_{i,\alpha}$ with
\begin{align}
\label{eq:divisorfib}
\begin{array}{cc}
 E_{i,\alpha} \rightarrow &D_{i,\alpha}  \\
& \downarrow \pi \\
 & \mathcal{C}_{\alpha} 
 \end{array} \, .
\end{align} 
Note that in general, there can be several divisors $D_{i,\alpha}$ labeled with $i=1\ldots n_{\text{nf},\alpha}$ that restrict to the same $\alpha=1\ldots n_{c}$ base curves $\mathcal{C}_\alpha$. 
This fibration structure allows to deduce the cohomology of \eqref{eq:divisorfib} from those of fiber and base using the Leray-Hirsch theorem \cite{Mavlyutov:2000dj,Greiner:2017ery}. Since the fibers $E_{i,\alpha}$ are compact and connected its Hodge numbers are trivial but $h^{0,0}(E_{i,\alpha}) = h^{2,2}(E_{i,\alpha})= 1$. This data is enough to show that the divisors $D_{i,\alpha}$ support a non-trivial one-form cohomology whose dimensions is given by
\begin{align}
\label{eq:hodge1}
 h^{1,0}(D_{i,\alpha})=  h^{1,0}(\mathcal{C}_{g_\alpha})    \cdot   h^{0,0}(E_{i,\alpha}) =   g_{\alpha}  \,  \text{ for } i=1 \ldots n_\alpha \, .
\end{align}   
This non-trivial one-form cohomology embeds non-trivially into the four-fold $X_4$ and is the main source for the three-form cohomology. This fact has been shown in  \cite{Mavlyutov:2000dj,Greiner:2017ery} via the 
 so called Gysin isomorphism $\iota$. For a smooth toric hypersurface $X_4$ and under the assumption that $D$ is toric, the Gysin spectral sequence has been evaluated in \cite{Mavlyutov:2000dj}. The result is that those divisors that admit a non-trivial one-form cohomology essentially inject a non-trivial three-form cohomology in $X_4$ of that same dimensions
\begin{align}
\label{eq:Gysin}
\oplus_{i,\alpha} H^{1,0}(D_{i,\alpha}) \xrightarrow{ \oplus_{i,\alpha} \iota:   }  H^{2,1}(X_4) \, .
\end{align}
Putting the pieces together we deduce the contribution of non-flat fibrations to the three-form cohomology as
\begin{align}
\label{eq:hodge}
h^{2,1}_\text{non-flat}(X_4) =  \sum_\alpha g_{\alpha}   \cdot n_{\alpha} \, .
\end{align}   
The Formula~\eqref{eq:hodge} can be used in general to also construct bases $B_3$ that exhibit themselves non-trivial three-forms $h^{2,1}(B_3)$ as has been done in \cite{Greiner:2017ery}. The trick is simply too look for bases that themselves admit divisors with non-trivial one-form cohomology. This however is not the focus of study for this work. Moreover, although eqn.~\eqref{eq:Gysin} has been proven in the framework of toric hypersurfaces, it is plausible to be of general validity. In the end from an intuitive point we might simply think of those three-forms to come from the sections of the genus g surfaces twisted by the surface component.

  As this work mainly focuses on the contribution of non-flat fibers, we will always consider bases with only  trivial fibration structure in \ref{sec:three} and \ref{sec:four}. Analogous to three-folds and also arguing from a physics perspective in Section~\ref{ssec:4612} each of those surface $E_{i,\alpha}$ contributes an additional K{\"a}hler parameter\footnote{The exact identification is given again via the Gysin spectral sequence \cite{Mavlyutov:2000dj} that reads $\oplus_{i,\alpha} H^{0,0}(D_{i,\alpha}) \xrightarrow{ \oplus_{i,\alpha} \iota:   }  H^{1,1}(X_4) \, .$. }.
We summarize the contributions of all non-flat surfaces using  \eqref{eq:hodge} as
\begin{align}
h^{1,1}(X_4)_\text{non-flat} =\sum_\alpha^{n_{c}} n_{nf,\alpha} \, , \qquad h^{2,1} (X_4)_\text{non-flat} = \sum_\alpha^{n_{c}} g_\alpha \cdot n_{\text{nf},\alpha} \, .
\end{align}  
We can now turn to  implications for the physics of F-theory compactifications.
The concrete examples presented in the sections below all make use of the toric description in terms of the Batyrev construction. For those exists a nice way to compute the Hodge numbers from the combinatorial data of the polytope $\Delta$ which we review in  Appendix~\ref{app:Batyrev}. This allows first to explicitly construct non-flat four-folds via toric geometry and to explicitly show the validity of the expressions above.
 \subsection{(4,6,12) and non-flat resolutions}
 \label{ssec:4612}
 The structures of non-flat fibers appear naturally when one considers elliptic fibrations that exhibit non-minimal singularities at codimension two (or higher). In the Weierstrass model this amounts to loci where the functions $f,g$ and discriminant $\Delta$ admits a vanishing order in the window\footnote{In codimension three \cite{Apruzzi:2018oge}, this window is bounded by the vanishing orders $(12,24,36)$ from above.} 
 \begin{align}
 \label{eq:nonflatwindow}
 (4,6,12) \leq \text{ord}_{\text{van}}(f,g,\Delta) < (8,12,24) \, .
 \end{align}
 In order to discuss those cases, we consider three-folds and the occurrence of those singularities in some detail first. 
 In a threefold $X_3$ these singularities occur over smooth points of the base $B_2$ and signal the presence of non-perturbative objects that become light in the six dimensional F-theory. The simplest example are those of E-string theories that exist at the lowest end of the window \eqref{eq:nonflatwindow}. These theories are typically engineered by collisions of $\mathfrak{e}_8$ and $\mathfrak{su}_1$ branes in a point of $B_2$.  
    In a threefold $X_3$, there are three possibilities to remove those singularities and make the geometry smooth\footnote{We assume a resolution of all codimension one $ADE$ type of singularities here.}:
 \begin{enumerate}
 \item Performing a complex structure deformation that removes the singularity. This deformation corresponds to the Higgs branch of the 6D strongly coupled theory.
 \item Blowing up the intersection point(s) in $B_2$ until the fiber becomes regular. This resolution corresponds to the tensor branch of the 6D strongly coupled theory. 
 \item Resolving the fiber in a non-flat way without changing the base. In M-theory this represents a point in the 5D  
   Coulomb branch of the circle reduced 6D SCFT, related via the usual F/M-theory duality.
 \end{enumerate}
 Note that the two different resolutions (2) and (3) should not be thought of largely different theories as they collapse down to the same singularity. Instead these resolutions should be thought of as different points 
  in the extended K{\"a}hler cone of the resolved threefold $X_3$ \cite{Apruzzi:2018nre,Apruzzi:2019opn,Apruzzi:2019vpe} related by flop transitions. Hence their  K{\"a}hler moduli spaces are the same and thus in 5D the dimension of the Coulomb branches are the same too. Since the 5D Coulomb branch is the sum of the 6D tensor branch plus 
the rank of the gauge algebra factors we can write
\begin{align}
\label{eq:nonflath11}
h^{1,1}_{\text{non-flat}}(X_3)= \text{dim}(\text{Coulomb}_{5D})=\text{dim}(\text{Tensor}_{6D})+\text{rank}(\text{G$_{6D}$}) \, .
\end{align}
For the simplest case of an E-string theory the $(4,6,12)$ collision can be avoided by a single blow-up in $B_2$ which yields dim(Tensor$_{6D}$)$=1$ and no gauge symmetry. In the same way the non-flat resolution can be performed by a single surface over that point with $h^{1,1}_{\text{non-flat}}=1$ as expected. Other non-trivial higher rank examples are discussed in Section~\ref{sec:four}.  

Having clarified the geometric and physics implications of non-flat fibrations in three-folds we can move to a four-fold.
The main point for geometry is that the non-flat resolution over a codimension two locus do not depend on whether it happens over points in a three-fold base $B_2$ or curves in a four-fold base $B_3$. The idea is analogous to the resolution of ADE singularities
at codimension one in an elliptic K3 and those at codimension one in an elliptic three-fold. The general resolution procedure is exactly the same and they contribute with the same amount of fibral divisors in the Shioda-Tate-Wazir decomposition eqn.~\ref{eq:stwX3}. An important difference  in a three-fold though is whether a fiber singularity is (semi-)split or non-split. Such effects are caused by some additional monodromy effects along the codimension one curve in $B_2$. Their presence affects the resolution of the fiber singularities  by identifying fibral curves and effectively folds the ADE singularity by an outer automorphism to a non-simply laced algebra. Similar to those, one might expect monodromies to also be present in four-folds that act non-trivial on the non-flat fiber. Such cases are not discussed in this work but left for future investigations. 
In the absence of such monodromies, the number of non-flat resolution surfaces in a four-fold can directly be inferred from those of three-folds $X_3$. From a physics perspective this makes sense as the theories might simply be viewed as compactifications of the 6D/(5D) theories on Riemann-surfaces $\mathcal{C}_{g_\alpha}$.\footnote{Cases with monodromy might directly incorporate possible twisted circle reductions \cite{Bhardwaj:2019fzv} in the geometry.}
 
This physics interpretation allows us to express the contributions of non-flat fibers to $h^{1,1}(X_4)$ and $h^{2,1}(X_4)$ in terms of 6D/5D SCFT data from eq.~ \eqref{eq:nonflath11}   and eq.~\eqref{eq:hodge} as
\begin{align}
\label{eq:HodgeNF4fold}
h^{1,1}_{\text{non-flat}}(X_4) =& \sum_\alpha^{n_c}  \text{dim}(\text{Coulomb}_{5D})_\alpha \, ,  \\
h^{2,1}_{\text{non-flat}}(X_4) =&  \sum_\alpha^{n_c}   \text{dim}(\text{Coulomb}_{5D})_\alpha g_\alpha \, , 
\end{align}
when summing over the $n_c$ curves $\mathcal{C}_\alpha$ in the base.
 In Section~\ref{sec:three} we   consider several  examples of four-folds with E-string curves as well as higher rank examples in Section~\ref{sec:four}.
 
  \subsection{Singlets in M/F-theory duality on elliptic four-folds}
 \label{ssec:HodgeNumbers}
In this section we review the contribution of three-form cohomology in the F/M-theory, following \cite{Grimm:2012yq}. In order to do so we consider M-theory on a four-fold $X_4$ and 
 consider the contributions that lead to neutral chiral fields in 4D.  
We first focus on those contributions that solely come from the base $B_3$ and lift to neutral singlets in the 4D F-theory. There are complex scalars $T_\kappa$ with $\beta=1\ldots h^{1,1}(B_3)+1$ that come from the expansion of the K{\"a}hler form $J $ and RR four-form $C_4$ of IIB string theory.  In a basis of $ \omega_\kappa \in H^{1,1}(B_3)$  we expand those as
\begin{align}
J = v ^{\kappa} \omega_\kappa \, , \qquad C_4 =\mathcal{B}_2^\kappa \wedge \omega_\kappa + \ldots \, .
\end{align}
Dualizing the two-forms $\mathcal{B}_2^\kappa$ in 4d gives rise to axions $\rho_\kappa$ that combine with $v^\kappa$ to complexified K{\"a}hler moduli $T_\kappa$. These axions can be gauged and are important for anomaly cancellation.  
Further chiral singlets originate from complex structure moduli $h^{3,1}(X_4)$ of the full four-fold. Our main interest though are the chiral singlets that are inherited from $h^{2,1}(X_4)$ that do not come from the base and in our case are given by
\begin{align}
h^{2,1}(Y_4) - h^{2,1}(B_3) = h^{2,1}(X_4)_{\text{non-flat}} \, .
\end{align}
These are obtained from the expansion of the M-theory  $C_3$-form and are denoted by $N^\beta$ with $\beta=1\ldots h^{2,1}(X_4)_{\text{non-flat}}$. These singlets are unfortunately not very well understood. I.e. it is not clear whether the axion parts  $\sigma$ of the $N^\beta$ couple to curvature terms of the form $\sigma F \wedge F$ and $\sigma R \wedge R$ in the 4D effective action. These singlets though appear in an interesting way \cite{Grimm:2010ks} in the K{\"a}hler moduli, given as
\begin{align}
T_\kappa = \frac12 w_{\kappa, \beta, \gamma} v^\beta v^\gamma + \frac14 d_{\kappa, \beta, \gamma} (N+ \overline{N})^\beta (N+\overline{N})^\gamma + i \rho_\kappa \, ,
\end{align}
with $w$ the base intersection form on $B_3$ and $d_{\kappa, \beta, \gamma}$ a holomorphic function on the complex structure moduli space. The curious observation though is, that the $N^\beta$ enjoys of an additional discrete symmetry $N\rightarrow -N$ that the other chiral singlet fields do not have. Understanding these fields and their couplings is beyond the scope of this note. Instead this review should serve as a motivation of why these singlets are interesting and that non-flat fibrations naturally produce them. 
 \subsection{Higgs branches and non-perturbative couplings}
 \label{ssec:HiggsCouplings}
Having discussed the role of the resolutions of non-minimal singularities in three-and four-folds in Section~\ref{ssec:4612} we now want to consider their deformations.   These deformations can be of very general type e.g. they can fully remove all singular fibers resulting in a broken gauge group. The specific kind of deformations we want to consider here are those which keep the gauge group $G$ but only create/remove $(4,6,12)$ loci at codimension two. For concreteness we fix such a locus to be the vanishing of the ideal $I_{\text{E-string}}=\{z,p \}$. For this we consider the inverse problem by starting with a three-fold geometry that does not posses any $(4,6,12)$ yet. This geometry should have some non-Abelian gauge algebra $G$ localized over $z=0$ and possibly additional Abelian factors. In general these gauge group factors are expected to lead to massless hypers in 6D that carry non-trivial representations $\mathbf{R}_i$. These matter representations are found at loci where the vanishing order of the discriminant $\Delta$ in the Weierstrass form enhances. Such loci we denote by the vanishing ideals
 $I_{\text{matter},i}$ which might be very complicated and not necessarily of complete intersection type. For simplicity we fix one of them to be $I_{\text{matter},0}=\{z,p\}$. At next we perform a complex structure deformation, such that one polynomial factors out one power of $z$ as
 \begin{align}
\label{eq:conifoldsplit}
 a  \rightarrow z\,  b  \, ,
\end{align} 
where $b$ itself is some other polynomial. For some of the matter loci $I_{\text{matter},i}$ we now require that these are themselves non-trivial in $\{ a,p\}$ which we denote as $I_{\text{matter},k}(a,z,p)$. Those ideals we require to become reducible upon the factorization eqn.~\eqref{eq:conifoldsplit} as
\begin{align}
\label{eq:csfactorization}
I_{\text{matter},k}(a,z,p) \xrightarrow{a  \rightarrow z\, b } I_{\text{matter},0} \oplus \hat{I}_{\text{matter},k} \, .
\end{align}
The deformation eqn.~\eqref{eq:conifoldsplit} forces  matter ideals to be moved onto the $I_{\text{matter},0}$ locus. Since the loci themselves are identified with matter loci, the Weierstrass coefficients  $(f,g,\Delta)$ admit a non-trivial vanishing order over them before tuning. The factorization in eqn.~\eqref{eq:csfactorization} therefore increases the vanishing order of the $I_{\text{matter},0}$ to the E-string loci $I_{\text{E-string}}$.
From the field theory side, such E-string transitions decrease the hypermultiplet sector $\mathcal{S}$ and turns them into non-perturbative E-string sector. The way we have set up those transitions above, might seem odd and artificial at first glance. However, those transitions are highly constrained by the 6D anomaly conditions due to the fact that all E-string points should admit a tensor branch where the field theory anomalies are properly canceled. Hence physics tells us that the 6D matter content must change in such a transition and hence the requirement \eqref{eq:csfactorization} is actually very natural to appear. In particular when the gauge group $G$ stays fixed such transitions are essentially unique \cite{Anderson:2015cqy,Dierigl:2018nlv}.  
      One obvious constrained is given by the gravitational anomaly which fixes the dimensions of  representations in $\mathcal{S}$ to $\sum_k \text{dim}(\mathbf{R}_k)=29$ as it is the same contribution of a 6D tensor multiplet. Hence the transition can never involve a representation $\mathbf{R}$ with dimension larger than $28$\footnote{If $\mathbf{R}$ is (pseudo)real this condition can be relaxed by a factor $1/2$. Moreover note that the gauge group $G$ must be subgroup of $E_8$  the flavor symmetry of the E-string theory.}.
Consistency of the physics and hence the geometry allows therefore to deduce  the subset of matter ideals $I_{\text{matter},k}$ in $I_{\text{matter},i}$ that admits the correct factorization properties as dictated by the 6D anomalies.

These structures generalize in a straight forward manner to four-folds. Indeed, the various polynomials important in the discussion can simply be taken to be sections on the base $B_3$. 
I.e. the same kind of complex structure deformation creates $(4,6,12)$ curves specified by the very same  codimension two ideals that lead to the factorization of matter curves $I_{\text{matter},k}$. However we can use this structure in order to make an additional observation in four-folds:
The factorization property \eqref{eq:csfactorization} under the deformation \eqref{eq:conifoldsplit} {\it guarantees }  
 the matter loci $I_{\text{matter},0}$ and $I_{\text{matter},k}(a,z,p)$ to all vanish at the {\it codimension three} point(s) $z=a=p=0$ {\it before} that transition. The prescribed deformation simply ensured that the polynomial $a$ vanishes to first order at  $z=p=0$ in order to obtain the $(4,6,12)$ loci. Hence even when the $(4,6,12)$ loci are absent at $z=p=0$, it is present at the codimension three locus  $a=z=p=0$. Indeed this effect is ensured by the reducibility of the matter ideals $I_{\text{matter},k}$ enforced by the  6D E-string transition. \\
As a result we see the $(4,6,12)$ curve in the four-fold base $B_3$ was simply deformed to one codimension lower where all curves of the ideals $I_{\text{matter},k}$ lie on. Enhanced singularities in the elliptic fiber at codimension three are generically interpreted as Yukawa couplings in the 4D superpotential $\mathcal{W}$. In the IIB picture these are induced from intersections of three matter curves in the respective point that is systematically been tracked by the F-theory torus. 
However when the fiber becomes of $(4,6,12)$ type one expects non-perturbative effects to be present similar to the E-string theories in 6D. Indeed codimension three points of such non-minimal type have been studied in \cite{Achmed-Zade:2018idx} where they were shown to lead to gauge invariant {\it four-point} couplings in the 4D superpotential. In the IIB picture these couplings are generated by D1 instanton strings that stretch between the involved matter matter curves that meet in the $(4,6,12)$ point.   

The main point of our geometric construction is, that it allows  to interpret those 
codimension three $(4,6,12)$ points naturally in terms of the 6D E-string transition. As the 6D anomalies have dictated the matter in the transition they also fix the very same curves in 4D that meet in the $(4,6,12)$ point at codimension three. These points and the matter curves $I_{\text{matter},k}$ that meet them are therefore naturally interpreted as the remnants of the 6D E-string transitions which lead to the non-perturbative four-point coupling of type
\begin{align}
\mathcal{W} \ni \prod^4_{i \in k} \mathbf{R}_k  \, ,
\end{align}
where representations $\mathbf{R}_k$ may occur multiple times. E-string transitions that preserve the total gauge group have been classified in \cite{Dierigl:2018nlv} including the respective change in the 6D matter spectrum. Hence knowledge can be used to infer the induced 4D non-perturbative couplings. 
A simple example is that of an $SU(7) \times U(1)$ 6D gauge theory. Anomalies of the 6D tensor branch force the following change in the hypermultiplet sector $\Delta \mathcal{S}=\left( \mathbf{21}_{-\frac{q}{3}} \oplus \mathbf{7}_{q} \oplus \mathbf{1}_0 \right)$ in an E-string transition. By the arguments above, the very same representations must be present in the four-point coupling that resembles the codimension three $(4,6,12)$ point in the four-fold. Indeed, the lowest order gauge invariant coupling \cite{Feger:2019tvk}  in the superpotential appears at fourth order as
\begin{align}
\mathcal{W} \ni \mathbf{21}_{-\frac{q}{3}} \cdot \mathbf{21}_{-\frac{q}{3}} \cdot  \mathbf{21}_{-\frac{q}{3}} \cdot \mathbf{7}_q \, .
\end{align}
In Section~\ref{sec:three} geometry and physics of similar examples are discussed in detail.  
\section{E-string transitions in three- and four-folds }
\label{sec:three} 
\begin{figure}[t!]
\begin{center}
	\begin{tikzpicture}[scale=1.3,
    ->,
    >=stealth',
    auto,node distance=3cm,
    thick,
    main node/.style={circle, draw, font=\sffamily\Large\bfseries}
    ]]
    
    	\node (A) at (-2.8,1) [ text width=3.5cm,align=center] {  $\Delta \chi \neq 0 $
		}; 
		
		 	\node (A) at (0.2,1) [ text width=3.5cm,align=center] {  $\Delta \chi = 0 $
		}; 
			\node (A) at (-2.8 ,1.5) [ text width=3.5cm,align=center] { (transition 1)
		}; 
				\node (A) at (0.2 ,1.5) [ text width=3.5cm,align=center] { (transition 2)
		}; 
		
	\node (A) at (-4.5,0.8) [ text width=3.5cm,align=center] {  $T^2$
		}; 
	\draw [  thick, right hook-latex ] (-4.5, 0.6) -- (-4.1 , 0.2);
	\node (A) at (-1.5,0.8) [ text width=3.5cm,align=center] {  $T^2$
		}; 
	\draw [  thick, right hook-latex ] (-1.5, 0.6) -- (-1.1 , 0.2);
	
		\node (A) at (1.5,0.8) [ text width=3.5cm,align=center] {  $T^2$
		}; 
	\draw [  thick, right hook-latex ] (1.5, 0.6) -- (1.9 , 0.2);

	\node (B) at (-4,0) [ text width=3.5cm,align=center] {  $X_{3,a}$
		}; 
		\draw[ thick, -> ] (-4,-0.2 ) -- (-4,-0.8);  
		\node (A) at (-4,-1) [ text width=3.5cm,align=center] {  $B_2$
		};

		\draw[ thick,right hook-latex  ] (-3.7,-0.2) -- (-2.5,-1.2);  
		\node (A1) at (-2.2,-1.3) [ text width=3.5cm,align=center] {  $X_{4,a}$
		}; 
		\draw[   -> ] (-2.2,-1.5 ) -- (-2.2,-2.1);  
		\node (A) at (-2.1,-2.3) [ text width=3.5cm,align=center] {  $\mathbb{P}^1$
		}; 
		
		\draw[ thick,right hook-latex  ] (-3.8,-1.2) -- (-3.2,-1.7);  
		
		\draw[   -> ] (-2.4,-1.4 ) -- (-2.9,-1.7);  
		
		\draw[   -> ] (-2.8,-1.9 ) -- (-2.4,-2.2);  
		
		\node (A) at (-3,-1.8 ) [ text width=3.5cm,align=center] {  $B_3$
		};

		\node (A) at (0.9,-2.3) [ text width=3.5cm,align=center] {  $\mathbb{P}^1$
		}; 
		
		\draw[ thick,right hook-latex  ] (-0.8,-1.2) -- (-0.2,-1.7);  
		
		\draw[   -> ] (0.6,-1.4 ) -- (0.1,-1.7);  
		
		\draw[   -> ] (0.2,-1.9 ) -- (0.6,-2.2);  
		
		\node (A) at (0,-1.8 ) [ text width=3.5cm,align=center] {  $B_3$
		}; 
		
			\draw[ thick, -> ] (-1,-0.2 ) -- (-1,-0.8);  
		\node (A) at (-1,-1) [ text width=3.5cm,align=center] {  $B_2$
		}; 
		
		
			\node (A) at (3.9,-2.3) [ text width=3.5cm,align=center] {  $\mathbb{P}^1$
		}; 
		
		\draw[ thick,right hook-latex  ] (2.2,-1.2) -- (2.8,-1.7);  
		
		\draw[   -> ] (3.6,-1.4 ) -- (3.1,-1.7);  
		
		\draw[   -> ] (3.2,-1.9 ) -- (3.6,-2.2);  
		
		\node (A) at (3,-1.8 ) [ text width=3.5cm,align=center] {  $ \widehat{B_3}$
		}; 
		
			\draw[ thick, -> ] (2,-0.2 ) -- (2,-0.8);  
		\node (A) at (2,-1) [ text width=3.5cm,align=center] {  $  \widehat{ B_2}$
		}; 
		 
		 \node (C) at (-1,0) [ text width=3.5cm,align=center] {  $X_{3,b}$};
		  \node (D) at (2,0) [ text width=3.5cm,align=center] {  $X_{3,c}$};

\draw[ thick,right hook-latex  ] (-0.7,-0.2) -- (0.6,-1.1);  
\draw[ thick,right hook-latex  ] (2.3,-0.2) -- (3.6,-1.1);  

\draw[ thick,right hook-latex  ] (-3.7,-0.2) -- (-2.5,-1.2);  
		\node (B1) at (0.8,-1.3) [ text width=3.5cm,align=center] {  $X_{4,b}$};	 
		\node (C1) at (3.8,-1.3) [ text width=3.5cm,align=center] {  $X_{4,c}$};	
  \path[every node/.style={font=\sffamily\small}]
 
   ;
  \draw
    (B) [out=30, in=150, blue] to  (C);
       \draw (C) [out=30, in=150, blue] to  (D);
 
	\node (A) at (0.1,-3) [draw,rounded corners,very thick,text width=3.4cm,align=center] {    Non-flat in codim 2 
		}; 
			\node (A) at (-3.5,-3) [draw,rounded corners,very thick,text width=3.4cm,align=center] {Non-flat in codim 3 
		}; 
			\node (A) at (3.2,-3) [draw,rounded corners,very thick,text width=2cm,align=center] {Flat 
		}; 
 
		\end{tikzpicture}
\end{center}
\caption{{\it \label{fig:fourfoldtransitionsFibered}Summary of a chain of conifold transitions in elliptic four-folds. 
The starting four-fold $X_{4,a}$ is constructed from a three-fold $X_{3,a}$ fibered over a $\mathbb{P}^1$. We then perform conifolds in four-folds inherited from the three-folds. Conifold 1 renders $X_{4,b}$ non-flat and conifold two removes the non-flat fiber again by a birational base change (partially). }}
 \end{figure}
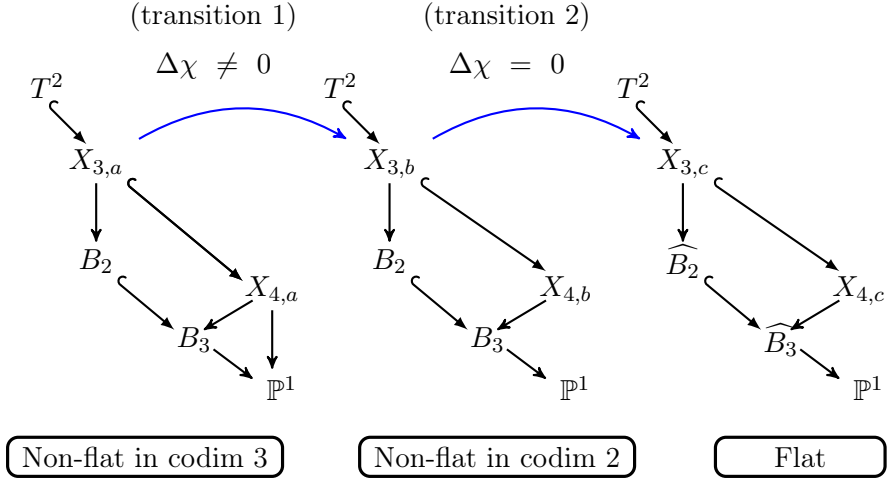

One way to obtain three-folds with non-flat fibers is to perform conifold transitions  \cite{Buchmuller:2017wpe,Dierigl:2018nlv} that originate  from those that are flat. We will adopt the very same strategy here for four-folds.
Concretely, we want to perform conifold transitions among three types of (compact) four-folds $X_{4,a}, X_{4,b}$ and $X_{4,c}$ as summarized in Figure~\ref{fig:fourfoldtransitionsFibered}. To stay close to the analogy of three-folds, we actually construct the four-folds as three-fold fibrations over another $\mathbb{P}^1$. This allows us to simply extract the three-fold conifold to four-folds. In this way, we are able to use the 6D anomalies of E-string transitions and to extend those to 4D. Note that this is just an auxiliary construction for illustrational purposes i.e. the logic also works for general four-folds.\\
 In order to investigate the non-flat fiber structure and its underlying physics, we perform conifold transitions that do not change the codimension one and Mordell-Weil structure i.e the 4D gauge group. We start in $X_{4,a}$ which admits a non-flat fiber in codimension three which is of course absent in the analogous three-fold $X_{3,a}$.   
    We then perform a complex structure deformation to enhance this non-flat fiber to curves in one codimension higher. Resolving the geometry fully leads to $X_{4,b}$.  These four-folds will exhibit in general a non-trivial three-form cohomology corresponding to the non-flat surface. In this section we consider E-strings curves only and hence the three-form contribution corresponds directly to the genus of the base curve. The final transition corresponds to a birational base change that removes all non-flat fibers. Keeping track of all Hodge numbers of both three and four-folds allows the computation of the Euler number too. Both in three-and four-folds we will observe how the first transition changes the Euler numbers while the second one does not. 
Our main example is going to be a  $(E_6 \times U(1))/\mathbb{Z}_3$  gauge theory. Here we focus on the details of the singular model and present details of the resolution in Appendix~\ref{app:E6toric}. Section~\ref{ssec:example2}  concerns similar configurations with gauge group $(SO(10) \times U(1)\times U(1))/\mathbb{Z}_4$. Besides being richer in structure which we present more technical details of the resolved geometry directly. All four-folds of consideration are compact and all Hodge numbers are computed via the Batyrev formulas.
\subsection{Example I: $(E_6 \times U(1))/\mathbb{Z}_3$}  
\label{ssec:example1}
\begin{figure}[t!]
 \begin{picture}(0,120)
 \put(0,20){\includegraphics[scale=1.5]{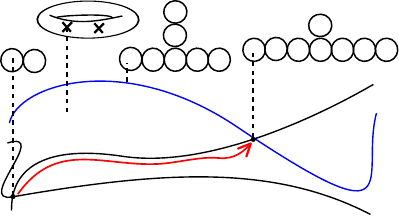}}
\put(230,20){\includegraphics[scale=2]{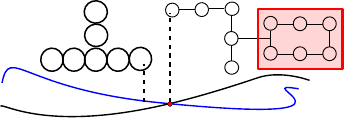}}
 \put(50,70){$ \mathfrak{e}_6$} 
  \put(30,60){$ \mathfrak{u}_1$} 
  \put(120,50){$ \mathbf{27}_{\frac13}$} 
   \put(7,15){$ \mathbf{1}_{-1}$} 
   \put(250,35){$ \mathfrak{e}_6$} 
   \put(320,15){E-string} 
 \end{picture}
 \caption{{\it \label{fig:3FoldE6}Two  $(E_6 \times U(1))/\mathbb{Z}_3$ geometries and the superconformal transition where an $\mathbf{1}_1$ singlet is tuned onto a $\mathbf{27}_{\frac13}$ locus which merges the local $\mathfrak{su}_2$ and $\mathfrak{e}_7$ enhancement into a non-flat fiber.}}
\end{figure}
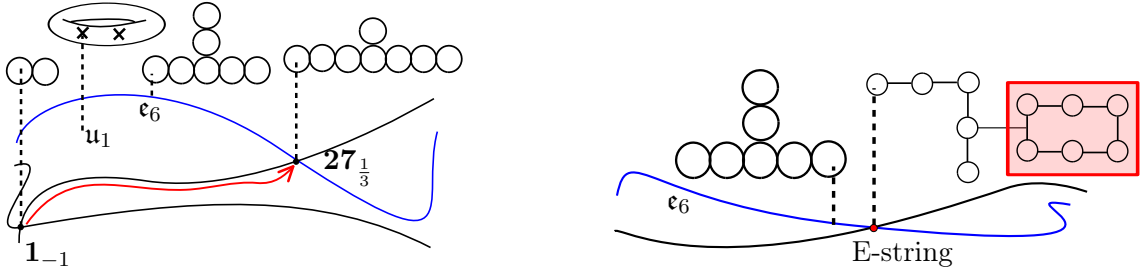

The starting theory we want to engineer is that of an $(E_6 \times U(1))/\mathbb{Z}_3$ 6D  gauge theory whose spectrum admits massless hypermultiplets in the representations 
\begin{align}
\mathbf{78}_0, \mathbf{27}_{\frac13} \, , \mathbf{1}_1 \, .
\end{align} 
In order to engineer this theory in F-theory we start with the Tate-model  
, given as  
\begin{align}
Y^2 = X^3 + a_1 X Y Z + a_2 X^2 Z^2 + a_3Y Z^3 + a_4X Z^4 + a_6 Z^6 \, .
\end{align}
The affine coordinates $[Y,X,Z]$ describe the elliptic fiber in an $\mathbb{P}^2_{3,2,1}$ ambient space.
The divisor $ Z=0$ is the zero-section and $a_i$  sections of line bundles of the base, fixed to be powers in the first Chern class of the base $[a_i] \in c_1(B)^i$.
The Weierstrass coefficients can easily be computed by completing the square and cube in $X$ and $Y$ which allows a mapping into the simpler Weierstrass form
\begin{align}
Y^2= X^3 + f X Z^4 + g Z^6 \, ,
\end{align}
where Weierstrass coefficients and discriminant is related to the Tate-form as
\begin{align}
\label{eq:wsf}
\begin{split}
f =& \frac{1}{48}(-(a_1^2 + 4 a_2)^2 + 24 (a_1 a_3 + 2 a_4))\, , \\
g  =& \frac{1}{864} ((a_1^2 + 4 a_2)^3 - 36 (a_1^2 + 4 a_2) (a_1 a_3 + 2 a_4) + 
     216 (a_3^2 + 4 a_6))\, , \\
     \Delta =& 4f^3+ 27g^2 \, .
     \end{split}
\end{align}
We start by engineering an $\mathfrak{u}_1$ gauge factor by setting $a_6=0$ globally which leads to an additional holomorphic section $s_1$ of the torus fiber at 
\begin{align}
s_1 : \{ X; Y ; Z \}=\{ 0 ;  -a_3  ; 1  \} \, ,
\end{align}
and hence a non-trivial Mordell-Weil group.
Matter that is charged under this gauge factor is found at the locus $a_3 = a_4=0$  where the fiber becomes of $I_2$ type. The $\mathfrak{e}_6$ factor is engineered over $\mathcal{Z}:\{ z=0\} $ according to the Tate-classification (e.g. see  \cite{Katz:2011qp}) by employing the factorization of the Tate-coefficients
\begin{align}
\{ a_1, a_2 ,a_3, a_4 \} \rightarrow \{ z a_{1,1} , z^2 a_{2,2}, z^2 a_{3,2}  , z^3 a_{4,3}  \} \, ,
 \end{align} 
which can be shown to lead to type $IV$ split fibers and hence an $\mathfrak{e}_6$ over $z=0$.  Matter in the $\mathbf{27}_{\frac13}$ representation\footnote{The Abelian charges are evaluated in the Appendix~\ref{app:E6toric}.} is found at the locus $z=a_{3,2}=0$. Up to now we have not specified the dimension of the base $B$ and hence the CY manifold. We start with a three-fold $X_{3,a}$ where the matter loci are just points that are counted by intersecting their divisor classes in the base cohomology. For some generic base\footnote{In order not to induce more singularities in the Weierstrass model, we restrict to bases where at most the divisor $\mathcal{Z}$ can admit negative self-intersections with $\mathcal{Z}^2 <-2$ but no other.} this leads to the multiplicities
\begin{align}
\label{eq:e6spectrum}
n_{\mathbf{27}_{\frac13}} = 6 (1-g) + \mathcal{Z}^2 \, , \qquad  n_{\mathbf{1}_1}= [a_{3,2}]\cdot [a_{4,3}]=34(g-1)-11 \mathcal{Z}^2 +12c_1^2 \, .
\end{align}
where $g$ denotes the genus of the curve $\mathcal{Z}$.
The resulting threefold geometry $X_{3,a}$ admits only minimal singularities. Hence the geometry $X_{3,a}$ can be fully resolved in a flat manner which we demonstrate in Appendix~\ref{app:E6toric}. An explicit geometry $Y_{3,a}$ with $\mathbb{F}_4$ base and   $\mathfrak{e}_6$ placed over the $-4$ curve is also given in Appendix~\ref{app:E6toric}. Using \eqref{eq:e6spectrum} one can compute the spectrum which admits two $\mathbf{27}_{\frac13}$-plets, 106 charged- and 163 uncharged singlets. The Hodge numbers of this geometry are given as
\begin{align}
\label{eq:HodgeX3a}
(h^{1,1},h^{2,1})_{\chi}(Y_{3,a})=(10,162)_{-304} \, .
\end{align}
The K{\"a}hler parameters are exactly as expected from the Shioda-Tate-Wazir theorem and the complex structure parameters precisely contribute the  amount of neutral singlets needed for anomaly cancellation. \\
Next we want to tune  E-strings along the $\mathfrak{e}_6$ gauge group by further performing the factorization 
\begin{align}
\label{eq:e6nffactor}
a_{4,3} \rightarrow z\,  b_{4,4} \, .
\end{align}
Counting the degrees of freedom we found this change to come at the cost of
\begin{align}
\Delta h^{2,1}(X)=(3 c_1-2\mathcal{Z})\mathcal{Z} = 6(1-g) + \mathcal{Z}^2 \, ,
\end{align}
complex structures in  $X_{3,b}$.
When plugging the factorization into \eqref{eq:wsf} one finds that one still obtains an $\mathfrak{e}_6$ type of gauge group over $\{ z=0\}$. However now the former matter locus $z=a_{3,2}=0$ got enhanced from a $(3,5,9)$ to a non-minimal $(4,6,12)$ singularity. Hence in $X_{3,b}$, the $\mathbf{27}_{\frac13}$-plets got exchanged to E-string theories.
  Note that the tuning also involved a change in classes of the $\mathbf{1}_1$ matter loci. 
The full charged matter spectrum of the 6D model is given as
\begin{align}
n_{\text{E-string}} = 6 (1-g) + \mathcal{Z}^2 \, , \quad n_{\mathbf{1}_1}=[a_{3,2}]\cdot [b_{4,4}] =40 (g-1) +12 c_1^2 -12 \mathcal{Z}^2 \, .
\end{align} 
We summarize the full change in the hypermultiplet sector $\mathcal{S}$ in the transition as
\begin{align}
\label{eq:deltaE6}
\Delta \mathcal{S} = n_{\text{E-string}} \cdot \left( \mathbf{27}_{\frac13} \oplus \mathbf{1}_1 \oplus   \mathbf{1}_0   \right)  \rightarrow n_{\text{E-string}}\cdot  [\text{E-string}] \, .
\end{align}
As already implicitly used in the considerations of anomalies before, a well defined SUGRA theory is only obtained when we get rid of those $n_{\text{E-string}}$ E-string contributions. This can be done blowing up the base exactly $n_{\text{E-string}}$ times, which brings us to geometry $X_{3,c}$. Since both theories $X_{3,a}$ and $X_{3,c}$ are consistent SUGRA theories one can analyze the possible representations that get lost in such an E-string transitions on very general grounds   \cite{Dierigl:2018nlv}. From the change of the 6D anomaly lattice one can derive the change of the spectrum in an $(E_6 \times U(1))\mathbb{Z}_3$ gauge group which in general must be given by the representations
\begin{align}
\label{eq:E6mttransition}
\Delta (\mathbf{27}_{q_{27}} + \mathbf{1}_{q_1} + \mathbf{1}_{0} ) \, , \qquad   q_{27}^2 = 9 q_1^2	 \, ,
\end{align}
as it is the case in the geometry above. Note that the transition requires a neutral singlet for each E-string it produces. This can be directly checked when considering the smooth non-flat resolution of $X_{3,b}$. The exact toric resolution is given in 
  Appendix~\ref{app:E6toric} as well es for the concrete three-fold $Y_{3,b}$ in detail. A schematic depiction of the fiber properties is shown in Figure~\ref{fig:3FoldE6}. There the non-flat surface component is highlighted as a red box, as well as its reducible components. In the example one observes these reducible components to arrange in a closed loop that is attached to other fibral divisors \cite{Apruzzi:2018nre,Apruzzi:2019opn,Apruzzi:2019vpe}.
Computing the Hodge numbers for $Y_{3,b}$ we find
\begin{align}
\label{eq:hodgenfe6}
(h^{1,1},h^{2,1})_{\chi}(Y_{3,b})=(12,160)_{-296} \, .
\end{align}
These Hodge numbers are consistent with the considerations made earlier. We remember that we have created two E-string type non-flat fibers that contribute one K{\"a}hler class each. This tuning though came at the cost of one complex structure for each of them.  To remove the non-flat fibers/E-string points we can blow-up the $-4$ curve which results in a $-6$ curve. In the toric setup we demonstrate this in Appendix~\ref{app:E6toric} for geometry $Y_{3,c}$ and show that it admits the same Hodge numbers  \eqref{eq:hodgenfe6} as three-fold $Y_{3,b}$.  
\subsection*{The four-fold compactification}
Having clarified the conifold transitions in the three-fold geometries and their associated F-theory physics, we can perform the same strategy in four-folds and consider their associated 4D physics. To keep matters close to 6D we construct the four-fold as a three-fold fibration over another $\mathbb{P}^1$.
In general we have lots of freedom to fiber $X_3$ over the new $\mathbb{P}^1$. For simplicity we chose this fibration to be trivial, which extends to the four-fold base as
\begin{align}
B_3 = B_2 \times \mathbb{P}^1 \, .
\end{align}
The matter loci of $X_{4,a}$ are readily obtained from $X_{3,a}$. These  codimension two loci become  
 Riemann-surfaces whose genus can be computed explicitly as
\begin{align}
g^\prime_{\mathbf{27}_{\frac13}} =1+24(1-g)+ 5 \mathcal{Z}^2 \, ,  \qquad 
g^\prime_{\mathbf{1}_{1}} =1+216 (3(g-1) +  c_1^2) - 203\mathcal{ Z}^2\, .
\end{align}
These quantities are expressed in terms of the classes the $B_2$ inherited classes $\mathcal{Z}$, $c_1$ and their genus $g$. Again the explicit resolved geometry $Y_{4,a}$ is given in Appendix~\ref{app:E6toric} for which we compute the Hodge numbers
\begin{align}
\label{eq:hodgenfe6}
(h^{1,1},h^{2,1},h^{3,1})_{\chi}(Y_{4,a})=(11,0,1447)_{8796} \, .
\end{align}.
 In comparison to $Y_{3,a}$ there is exactly one more K{\"a}hler class of the additional base $\mathbb{P}^1$.
Contributions to $h^{2,1}(Y_{4,a})$ are trivial, as the fibration is flat and the base with  $h^{2,1}(F_4 \times \mathbb{P}^1)=0$ does not contribute either. When moving to the matter sector we find the two $\mathbf{27}_{\frac13}$ plets in 6D to be localized over the genus $g^\prime=5$ curve in the $B_3$ base and similarly the singlets over a curve of genus $g^\prime_{\mathbf{1}_1}=1893$.

\begin{figure}[t!]
 \begin{picture}(0,130)
 \put(0,20){\includegraphics[scale=1.5]{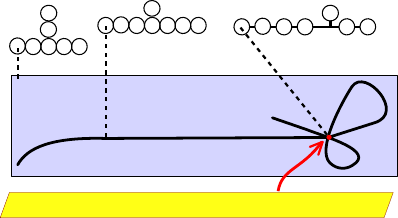}}
 \put(5,70){$\mathfrak{e}_6$}
  \put(50,60){$\mathbf{27}$}
  \put(155,50){$(\mathbf{27})^3$}
    \put(90,20){$U(1)$}
     \put(240,50){$\mathbf{27}_{\frac13}$}
     \put(260,28){$\mathbf{1}_{-1}$}
        \put(380,40){$(\mathbf{27}_{\frac13})^3  \mathbf{1}_{-1}$}
\put(230,20){\includegraphics[scale=2]{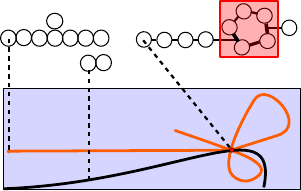}} 
 \end{picture}
 \caption{{\it \label{fig:4FoldE61} Two depictions of the $\mathfrak{e}_6$ divisor in blue and its fiber structure in the four-fold $X_{4,0}$ and $X_{4,a}$. In $X_{4,0}$ the $\mathbf{27}^3$ Yukawa point leads to a (reduced) $E_8$ fiber bouquet.
Upon adding a $\mathfrak{u}_1$ brane the $E_8$ fiber becomes non-flat fiber and leads to a four-point coupling. }}
\end{figure}
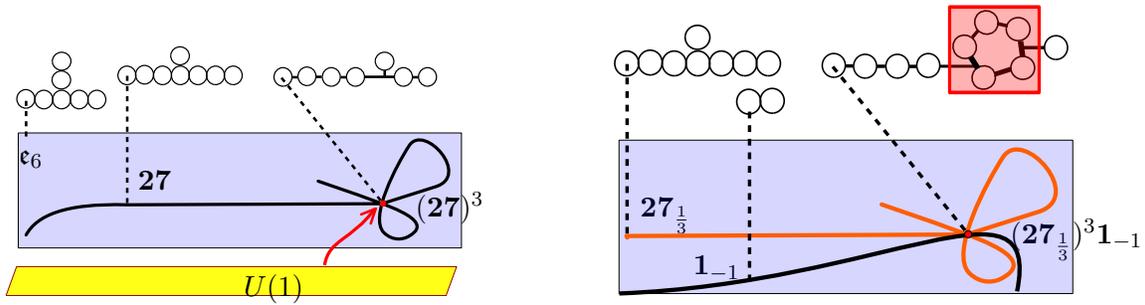

Before performing the first conifold transition, we consider interaction terms between the charged fields present in the 4D superpotential $\mathcal{W}$. 
A key feature of F-theory which makes it also relevant for phenomenological model building  is the presence of Yukawa couplings that are suppressed in perturbative IIB string theory \cite{Donagi:2008ca,Beasley:2008dc,Beasley:2008kw,Font:2012wq,Font:2013ida,Marchesano:2015dfa}. An example is that of the top quark in an $\mathfrak{su}_5$ theory. This coupling is schematically given by couplings of type  $\mathbf{10} \cdot \mathbf{10} \cdot \mathbf{5}$. In F-theory these points lead to 
 type $IV^*$ fiber degenerations. Those degenerations highlight local IIB string couplings of order one and hence strongly coupled in nature. Similarly, when one further unhiggses the $\mathfrak{su}_5$ to $\mathfrak{e}_6$ there are couplings of type
    $\mathbf{27} \cdot \mathbf{27} \cdot \mathbf{27}$. Over these points, the fiber enhances to a type $II^*$ singularity, i.e. a point of $\mathfrak{e}_8$.\footnote{An idea to infer the above coupling is by viewing the local enhancement as that of $\mathfrak{e}_8$ and decompose the fundamentals representation into those of  $\mathfrak{e}_6$ to infer the Yukawa coupling $\mathbf{248}^3 \ni \mathbf{27}^3$ .} 
        This fact can be readily checked when adding the term $Z^6 z^5 a_{6,5}$ to the Tate-model which breaks the $\mathfrak{u}_1$ gauge symmetry. In the following we will call this four-fold with only $E_6$ gauge group, $X_{4,0}$. For a fully resolved model we have depicted fiber structure of the $\mathfrak{e}_8$ point in Figure~\ref{fig:4FoldE61}. Note that the exact fiber topology in a smooth model might not be of precise $\mathfrak{e}_8$ shape. Instead the fiber structure might only be a bouquet which misses a couple of $\mathfrak{e}_8$ nodes \cite{Esole:2011sm}. This effect however does not to obstruct  the phenomenological implications of the Yukawa couplings in \cite{Marsano:2011hv,Collinucci:2016hgh}.\footnote{The effect on the fibers can be explained as an IIB orientifold effect that removes certain modes that correspond to the respective nodes in the diagram.}
    
By setting the polynomial $a_{6,5}$ in the Tate-model to zero restores the global $\mathfrak{u}_1$ gauge factor under which the   $\mathbf{27}$-plets are non-trivially charged. From a field theory perspective one therefore expects the   $\mathbf{27}^3$ Yukawa coupling to be forbidden. Therefore one might come to the conclusion that those intersection points should be absent in a given geometry. However, what we actually observe is, that the $\mathfrak{e}_8$ Yukawa points enhance to $(4,6,12)$ as discussed in Section~\ref{ssec:4612}.  
From the general discussion of Section~\ref{ssec:4612} we therefore a four-point coupling that involves matter representation of the 6D tensor transition eqn.~\eqref{eq:deltaE6}. 
Indeed in our situation the (4,6,12) singularity precisely occurs over the points $z=a_{3,2}=a_{4,3}=0$ where  $\mathbf{27}_{\frac13}$ and $\mathbf{1}_{-1}$ matter curves meet \footnote{This exact example has been anticipated in \cite{Achmed-Zade:2018idx}.}. The resulting four-point coupling is therefore schematically given as
\begin{align}
\mathcal{W} \ni \mathbf{27}_{\frac13} \cdot \mathbf{27}_{\frac13} \cdot \mathbf{27}_{\frac13}\cdot \mathbf{1}_{-1} \, .
\end{align}
From the perspective of the global tuning, this point is very analogous to an E-string transition in which we have tuned an extra $U(1)$ onto the $\mathfrak{e}_8$ Yukawa point as shown in Figure~\ref{fig:4FoldE61}.  Note also that generically those points come with a multiplicity when counting the intersection points as
\begin{align}
N_{\text{Yukawa}}(X_{4,a})= \mathcal{Z} \cdot [a_{3,2}]\cdot [a_{4,3}]= 2(48(1-g)-7 \mathcal{Z}^2  ) \, ,
\end{align}
that is $N_{\text{Yukawa}}(Y_{4,a})=152$ respectively. Finally we comment on the fiber structure of the resolved model $X_{4,a}$. Here the non-flat Yukawa point is depicted in Figure~\ref{fig:4FoldE61}. 
As opposed to the codimension one non-flat fibers, these do not contribute additional K{\"a}hler parameters as they simply come from the degeneration of one of the exceptional $\mathfrak{e}_6$ resolution divisors.
\subsubsection*{Enhancing to non-flat fibers in codimension two}
In the next step we want to perform the matter-transition we did in the three-fold given by \eqref{eq:e6nffactor} now in the four-fold geometry $X_{4,b}$. From the four-fold perspective, this tuning pushes the codimension three (4,6,12) singularity one codimension higher and enhances   the $\mathbf{27}_{\frac13}$-plet curve into that of an E-string.   On the other hand, from a 6D perspective  we can interpret this curve as the compactification of  
    $6(g-1)+\mathcal{Z}^2$ E-strings points on a $\mathbb{P}^1$ which makes them a single curve of genus
  $g^\prime_{\text{E-string}} =1+24(1-g)+ 5 \mathcal{Z}^2$. Similarly to the three-fold $X_{3,b}$ in the four-fold $X_{4,b}$ this E-string curve by obtaining contributions of the $\mathbf{1}_{1}$ singlet curve. Indeed,
    the singlet curve changes it genus by  $\Delta g_{\mathbf{1}_1} = 9 (8( - 8) g - \mathcal{Z}^2)$ to 
\begin{align}
\tilde{g_{\mathbf{1}_1}} = 1+ 4(180(g-1)+54c_1^2 -53 \mathcal{Z}^2) \, .
\end{align}
In the concrete example of four-fold $Y_{4,b}$ that is detailed in Appendix~\ref{app:E6toric} we compute the Hodge numbers as
\begin{align}
(h^{1,1},h^{2,1},h^{3,1})_{\chi}(Y_{4,b})=(12,5,1281)_{7776}  \, .
\end{align}
These Hodge numbers are to be understood from the general discussion in Section~\ref{sec:two}: First there is the E-string theory whose non-flat surface which contributes a single new K{\"a}hler class as its 6D rank is one. Note that we have started with two $6D$ E-string theories, that both contributed an individual K{\"a}hler class. Since both contributions are merged along the same genus-five surface in the base, they now contribute a single K{\"a}hler class.
  Second there is the contribution of the five $h^{2,1}$ three-forms. These are inherited from the rank one theory
  multiplied by the genus of its curve. 
 
Finally, we can discuss the removal of these non-flat surfaces by changing the  $B_3$ base analogous to the 6D case. We   take two {\it equivalent } ways by either blowing up the base $B_2$ to get $X_{3,c}$ and then compactify this configuration over another $\mathbb{P}^1$ or by blowing up the four-fold base in $X_{4,b}$ in the $B_2$ direction to get $X_{4,c}$. In both cases we end up with a base $\widehat{B}_3 = (BL B_2)\times \mathbb{P}^1$. Notably $X_{3,c}$ admits no E-string points anymore which might not the case for the four-fold $X_{4,c}$. This can be seen from realizing that the E-string curve class in the four-fold is given by
 \begin{align}
\mathcal{Z} \cdot [a_{3,2}]= 3 c_1(B_2)  \cdot  \mathcal{Z} - 2  \mathcal{Z}^2+ 6 H_1  \cdot \mathcal{Z}  \, .
\end{align}
When we pullback $ 3c_1(B_2) \cdot \mathcal{Z} \sim 2  \mathcal{Z}^2$ from $B_2$ this is enough to require absence of E-string points in a three-fold but not sufficient for the four-fold anymore. In order to do so we also need to require the relation $H_1 \cdot \mathcal{Z} \sim 0$ by e.g. perform additional blow-ups in the base.
Hence  even when started with a non-higgsable cluster in  6D, i.e. $E_6$ on a $-6$ curve that admits neither matter and is fully flat, the new $\mathbb{P}^1$ direction introduce them. The same is true for the explicit non-flat resolution in $Y_{4,c}$ where we have performed a base change solely in the $B_2$ direction. Performing this resolution has split the respective genus five in $B_3$ into six  genus-zero curves as depicted in Figure~\ref{fig:4FoldE62}.  The full four-fold geometry is given in Appendix~\ref{app:E6toric} and admits the Hodge numbers  
\begin{align}
\label{eq:hodgex4c}
(h^{1,1},h^{2,1},h^{3,1})_{\chi}(Y_{4,c})=(19,0,1269)_{7776} \, .
\end{align}
\begin{figure}[t!]
 \begin{picture}(0,130)
 \put(0,20){\includegraphics[scale=2]{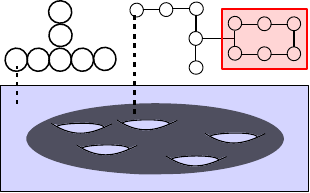}}
 \put(5,70){$\mathfrak{e}_6$}
  \put(80,73){E-string}
   \put(230,45){$\mathfrak{e}_6$}
  \put(380,60){E-string} 
\put(230,20){\includegraphics[scale=2]{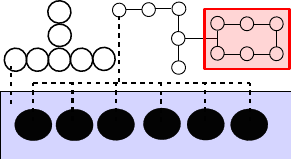}} 
 \end{picture}
  \caption{{\it \label{fig:4FoldE62} Depiction of the fiber structure of the $\mathfrak{e}_6$ divisor in blue. The left shows an E-string fiber over a genus five curve resolved as a non-flat fiber. On the right a base change has deformed this curve into  six genus zero curves that are removed by additional blow-ups. All transitions change the Hodge numbers but not the Euler number of the compact four-fold.}}
\end{figure}
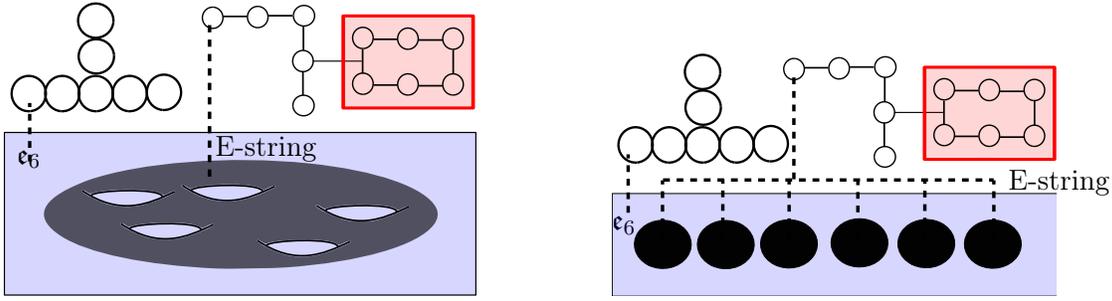
The Hodge numbers make sense as we have introduced two more resolution divisors to $B_2$. This change in the base has split up the single non-flat surface into six disconnected genus zero curves. As all of them host a non-flat fiber, they all contribute additional K{\"a}hler parameters to $Y_{4,c}$.  None of the non-flat surfaces contribute to $h^{2,1}(Y_{4,c})$  anymore. In order to obtain the fully flat four-fold, six more base blowups are required to split to remove all genus-zero curves. 
The exact polytope of $Y'_{4,c}$ is again given in the appendix and it is fully flat. Moreover it admits {\it exactly } the same Hodge numbers as $Y_{4,c}$ given in \eqref{eq:hodgex4c}.

Finally we are in the position to compare the transition between three-folds and those between four-folds. While removing non-flat fibers of $Y_{3,b}$ via a base change to $Y_{3,c}$ has not changed the Hodge numbers at all, this was not the case in the analogous transition of  four-folds. Here in fact virtually all Hodge numbers where involved in the transition.
 However an important observations is that all those transitions left the Euler number invariant. This observation is particularly important when it comes to the inclusion of $G_4$ flux which we have not addressed in this work. The relevant point is that it makes it easier to satisfy the change in the tadpole cancellation conditions eqn.~\eqref{eq:tadpolechange} when a consistent $G_4$ flux configuration has been found in either phase. We will comment on this point again in the conclusion. 
\subsection{Example II: $(SO(10) \times U(1)^2)/\mathbb{Z}_4 $}
\label{ssec:example2}
\begin{figure}[t!]
 \begin{picture}(0,136)
 \put(0,20){\includegraphics[scale=2]{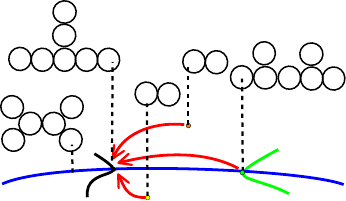}}
\put(230,20){\includegraphics[scale=2.3]{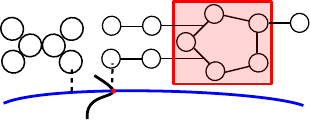}}
 \put(10,20){$ \mathfrak{so}_{10}$} 
  \put(40,10){$ \mathbf{16}_{-\frac14,-\frac12}$} 
  \put(130,20){$ \mathbf{10}_{\frac12,0}$} 
  \put(90,20){$ \mathbf{1}_{1,1}$} 
   \put(110,60){$ \mathbf{1}_{0,1}$} 
   \put(250,25){$ \mathfrak{so}_{10}$} 
   \put(310,25){E-string} 
 \end{picture}
 \caption{{\it \label{fig:3Foldso10}The fiber structure of the  $(SO(10) \times U(1)^2)/\mathbb{Z}_4$ model. On the left a complex structure deformation tunes four matter representations onto the same point while keeping codimension one fibers inert. Over this point the resulting fiber becomes non-flat which is shown via the red surface on the right.}}
\end{figure}
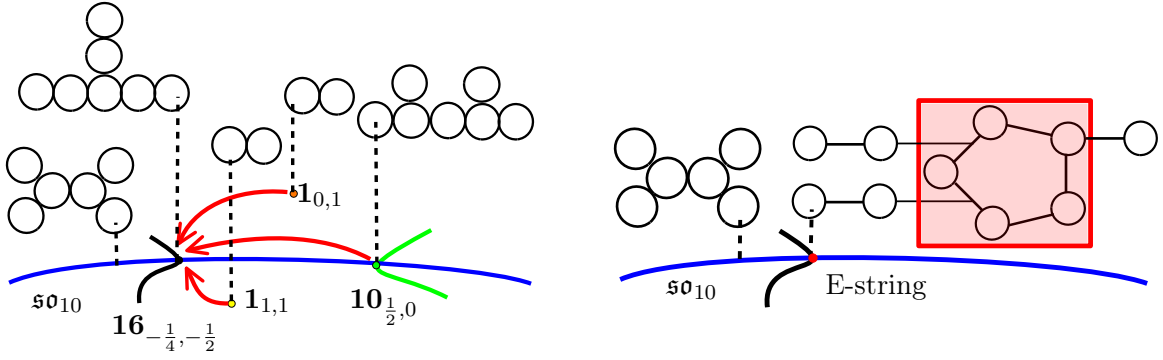

The second example works similar as the one before but admits a more rich matter structure. Moreover we use the chance to directly give more attention to the explicit toric resolution of the geometry.  Again we use  E-string transitions in 6D to understand four-point Yukawa couplings and the general change of the matter spectrum in 4D. The model at hand admits two $\mathfrak{u}_1$ gauge factors that are engineered with the most general elliptic model with three sections \cite{Cvetic:2013nia}. This model  is given by the elliptic curve in an $dP_2$  which we parametrize by the coordinates $\{ u,v,w\}$ and the two exceptional coordinates $e_1, e_2$.
  An additional $\mathfrak{so}_{10}$ singularity over $\mathcal{Z}: x_0=0$ can be engineered via a toric top \cite{Bouchard:2003bu,Buchmuller:2017wpe} which we resolve directly via the exceptional coordinates $f_2,f_3,f_4,g_1,g_2$ and $f_0$ being the affine node. The resolved hypersurface\footnote{One can return to the singular model by setting all exceptional coordinates to one and replaces $f_0$ with $x_0$.} is given as  
\begin{align}
\label{eq:so10a}
p&=  d_1 e_1^2 e_2^2 f_0 f_2^2 f_4 g_1 u^3 + 
 d_2 e_1 e_2^2 f_0^2 f_2^2 f_3 f_4 g_1^2 g_2 u^2 v + 
 d_3 e_2^2 f_0^2 f_2 f_3 g_1 u v^2 + d_5 e_1^2 e_2 f_2 f_4 u^2 w \nonumber \\ & + 
 d_6 e_1 e_2 f_0 f_2 f_3 f_4 g_1 g_2 u v w + d_7 e_2 f_0 f_3 v^2 w + 
 d_8 e_1^2 f_2 f_3 f_4^2 g_1 g_2^2 u w^2 + d_9 e_1 f_3 f_4 g_2 v w^2 \, .
\end{align}
The $d_i$ are to be interpreted as generalized Tate-coefficients that are sections of the base. This form of the elliptic curve can be mapped into Weierstrass form by blowing down all exceptional divisors and use the Arten-Tate algorithm. The Stanley-Reisner ideal for the chosen fiber triangulation is given by  
\begin{align}
\mathcal{SRI:} \{& e_2 w, e_1 e_2, e_2 v, e_2 f_0, e_2 f_3, e_2 g_1, e_2 g_2, u w, u v, f_0 u, f_3 u,  
f_4 u, g_1 u, g_2 u, f_2 w, f_2 v, f_2 f_3,\nonumber \\ & f_2 g_2, f_4 w, e_1 f_4, f_4 v, f_0 f_4, 
f_3 f_4, e_1 w, f_0 w, g_1 w, g_1 v, g_2 v, e_1 f_3, e_1 g_2, f_0 g_2, e_1 g_1  \} \, .
\end{align}
This model has been analyzed in detail in \cite{Buchmuller:2017wpe} where more details of the exact spectrum computation are given. We choose a simple two-fold base as $ B_2=\mathbb{F}_0$ and a fibration with bundle choice of the base sections $d_i$:
\begin{align}
\label{eq:so10classes}
\begin{array}{ll}
\mathcal{Z}: (D_{x_0}): H_1 \, , & [d_6]: H_1+2H_2 \, , \\
   \left[ d_1 \right]:H_1+2 H_2 \, , & [d_7]: H_1 + 2 H_2 \, ,\\
\left[d_2 \right]: 2H_2 \, ,& [d_8]: 2H_1 +2  H_2 \, ,\\
\left[d_3\right]: 2 H_2 \, ,& [d_9]: 2 H_1 + 2H_2 \, ,\\
 \left[d_5 \right]: 2H_1+2H_2 \, , & 
\end{array}
\end{align} 
where again $\mathcal{Z}$ is the $\mathfrak{so}_{10}$ divisor and $H_1$ and $H_2$ the two classes of  $\mathbb{F}_0$. 
With the help of the Weierstrass model one can find the reducible fiber components which give rise to the 6D matter. In Table~\ref{tab:SO10Spectrum} we have summarized multiplicity and matter representations.
\begin{table}
\begin{center}$
\begin{array}{|c | c|c|}\cline{2-3}
\multicolumn{1}{c|}{}&G & SO(10) \times U(1)^2 \\ \multicolumn{1}{c|}{}
&T & 1 \\ 
\multicolumn{1}{c|}{}& \mathbf{1}_0 & 43 \\ \hline
I_1: \quad  x_0=d_5=0 & \mathbf{16}_{-1/4, -1/2}:  &  2 \\
I_2: \quad x_0=d_7=0& \mathbf{16}_{-1/4, 1/2}:  & 2 \\
I_3: \quad x_0 = (d_3 d_5-d_1 d_7)=0 &\mathbf{10}_{1/2, 0}:  & 4  \\
I_4: \quad x_0=d_9=0 &\mathbf{10}_{1/2, 1}:  & 2  \\ \hline
I_5: \quad d_3 = d_7 = 0 & 1_{1,-1} & 6\\
I_6: \quad d_8 = d_9 = 0 & \mathbf{1}_{-1,-2} &8  \\
I_7: \quad d_9 = d_7 = 0 & \mathbf{1}_{0,2} &6  \\ \hline
I_8/ \{I_2,I_4,I_5,I_6,I_7 \}:  & \mathbf{1}_{0,1} &56 \\
I_9:  & \mathbf{1}_{1,0} &22 \\
I_{10}/ \{ I_4,I_6\}: & \mathbf{1}_{-1,-1} &26 \\ \hline
\end{array}$  \end{center}
\caption{{\it \label{tab:SO10Spectrum}Summary of the 6D matter spectrum. We also give the respective codimension two loci and their multiplicities using eqn.~\eqref{eq:so10classes}. The longer (quotient) ideals $I_{8,9,10}$ are given in eqn.~\eqref{eq:Iimatterloci}. }}
\end{table}
$I_{i}$ singlet loci that are given as: 
\begin{align}
\begin{split}
\label{eq:Iimatterloci} 
I_8:& \{ (d_5 d_7^3 d_8 d_9^2 - d_5 d_6 d_7^2 d_9^3 + d_3 d_5 d_7 d_9^4 + d_1 d_7^2 d_9^4 +
    d_7^4 d_8^3 x_0 - 2 d_6 d_7^3 d_8^2 d_9 x_0\\ & + d_6^2 d_7^2 d_8 d_9^2 x_0 + 
   2 d_3 d_7^2 d_8^2 d_9^2 x_0 - 2 d_3 d_6 d_7 d_8 d_9^3 x_0 + 
   d_3^2 d_8 d_9^4 x_0),  \\ & (-d_5 d_7^3 d_9^2 - d_7^4 d_8^2 x_0 + 
   d_6 d_7^3 d_8 d_9 x_0 - d_3 d_6 d_7 d_9^3 x_0 + d_2 d_7^2 d_9^3 x_0 + 
   d_3^2 d_9^4 x_0) \}\, ,   \\ I_9: &
  \{ (-d_3 d_6 d_7 + d_2 d_7^2 + 
  d_3^2 d_9)  , -  (-d_3 d_5 d_7 + d_1 d_7^2 + d_3^2 d_8 x_0) \} \, , \\ I_{10}: & 
  \{ (d_1 d_9^2 + d_3 d_8^2 x_0 - d_2 d_8 d_9 x_0), 
 d_5 d_9^2 + d_7 d_8^2 x_0 - d_6 d_8 d_9 x_0 \} \, ,
 \end{split}
\end{align}
The multiplicities can be computed by using the choices of classes \eqref{eq:so10classes} and the basic intersections on $\mathbb{F}_0$,  $H_1 \cdot H_2 =1$ and $H_i^2 = 0$.\footnote{For the quotient ideals, the contained loci have to be subtracted with multiplicities that are determined using the resultant.} This is enough to show 6D gauge anomaly cancellation. The toric realization of this model is given via the polytope
\begin{align}
\label{eq:so10polytope}
\begin{tabular}{ccc}
\begin{tabular}{|c|c|}\hline
\multicolumn{2}{|c|}{Generic Fiber}\\ \hline
e$_2$& (1,0,0,0) \\
w & (0,1,0,0) \\
u & (1,-1,0,0) \\
e$_1$&(0,-1,0,0)\\
v&(-1,0,0,0) \\ \hline 
\end{tabular}
&
\begin{tabular}{|c|c|}\hline
\multicolumn{2}{|c|}{$\mathbb{F}_0$ base}\\ \hline 
y$_1$ & (0,0,0,1) \\
 y$_0$ & (0,0,0,-1)\\
x$_1$ & (0,0,-1,0)\\ \hline
\end{tabular}
&
\begin{tabular}{|c|c|}\hline
\multicolumn{2}{|c|}{$\mathfrak{so}_{10}$ top}\\ \hline 
f$_0$ & (0,0,1,0) \\
f$_2$ & (1,0,1,0) \\
f$_3$&(0,1,1,0)\\
f$_4$&(1,1,1,0) \\  
g$_1$ & (1,1,2,0) \\ 
g$_2$ & (1,2,2,0) \\ \hline
\end{tabular}  
\end{tabular}\, ,
\begin{array}{l}
\Delta =\{\text{Fiber, Base, Top }\}: \\
(h^{1,1}, h^{2,1})_\chi (Y_{3,a})=(10,42)_{-64} \, .
\end{array}
\end{align}
Note that the elliptic fibration structure is inherited from a $dP_2$ fibration of the ambient spaces. This allows  also to directly obtain the projection $\pi$  from the ambient variety given via a projection onto the two last coordinates in $\Delta$.
   Using the $42+1$ neutral singlets also allows to show cancellation of the gravitational anomaly.
Next we want to perform an E-string transition which we do by factorizing another power of the $\mathfrak{so}_{10}$ divisor $x_0$ out of the base section $d_1$. This tuning effectively merges two $\mathbf{10}_{1/2,0}$-plets and two $\mathbf{16}_{-1/4,-1/2}$ as well as two $\mathbf{1}_{1,1}$ and $\mathbf{1}_{0,1}$-plets as one can explicitly see from Table~\eqref{tab:SO10Spectrum} and eqn.~\eqref{eq:Iimatterloci}. After this transition we have two non-minimal singularities over $d_5=x_0=0$. Lets again summarize the total change in the spectrum  $\mathcal{S}$
 \begin{align}
 \label{eq:so10transition}
 \Delta \mathcal{S}= 2\cdot \left( \mathbf{16}_{-1/4,-1/2} \oplus \mathbf{10}_{1/2,0} \oplus \mathbf{1}_{-1,-1} \oplus \mathbf{1}_{0,1} \oplus \mathbf{1}_{0,0} \right) \rightarrow 2  \cdot [\text{E-strings}]\, .
 \end{align}
This transition is fully consistent with the expectation of a tensor-matter transitions derived in \cite{Dierigl:2018nlv} 
The geometry that admits the two $(4,6,12)$ points can be resolved using toric geometry. This is done by simply adding the vertex  vertex $f_1 = (0,2,1,0)$ to the polytope \eqref{eq:so10polytope}. The addition of the vertex respects the fibration and does not change the base while also leaving the new polytope reflexive. Hence the the anti-canonical hypersurface is still Calabi-Yau, which we denote by $Y_{3,b}$ with Hodge numbers:
 \begin{align}
(h^{1,1},h^{2,1})_\chi (Y_{3,b}) = (12,40)_{-72} \, .
 \end{align}
 In comparison to $Y_{3,a}$ we find the expected loss of two complex structure parameters that are traded for the two K{\"a}hler parameters that come fro the non-flat fibers. In the toric realization, those are given by the ambient divisor   $f_3=0$ that intersect the base twice. 
 This can be seen from the new hypersurface which is explicitly given as  
\begin{align*}
p &=  d_1 e_1^2 e_2^2 f_0^2 f_1^3 f_3 f_4^2 g_1^3 g_2^2 u^3 + 
 d_2 e_1 e_2^2 f_0^2 f_1^2 f_3 f_4 g_1^2 g_2 u^2 v + 
 d_3 e_2^2 f_0^2 f_1 f_3 g_1 u v^2 + d_5 e_1^2 e_2 f_1 f_4 u^2 w \nonumber \\ & + 
 d_6 e_1 e_2 f_0 f_1 f_2 f_3 f_4 g_1 g_2 u v w + d_7 e_2 f_0 f_2 f_3 v^2 w + 
 d_8 e_1^2 f_1 f_2^2 f_3 f_4^2 g_1 g_2^2 u w^2 + d_9 e_1 f_2^2 f_3 f_4 g_2 v w^2 \, .
\end{align*}
A comment from the perspective of the toric geometry is in order. Here the generic fiber structure parametrized by  $u,v,w,e_1,e_2$ has not changed at all but the modification appears only at codimension two over $d_5=0$. There the  $f_3$ coordinate factors out globally and contributes a non-flat surface. Intersections of the non-flat surface can be computed via a triangulation that leads to the Stanley-Reisner ideal
\begin{align}
SRI: \{  &e_2 w, e_1 e_2, e_2 v, e_2 f_0, e_2 f_2, e_2 f_4, e_2 g_1, u w, u v, f_0 u, f_2 u, 
 f_3 u, f_4 u, g_1 u,   f_1 w, f_1 v, f_1 f_2, \nonumber \\ & f_1 g_1,  f_3 w, 
 e_1 f_3, f_3 v, f_0 f_3, f_2 f_3, e_1 w, f_0 w, f_4 w, f_4 v, g_1 v, e_1 f_2, 
 e_1 g_1, f_0 g_1, e_1 f_4  \} \, .
\end{align} 
The tuning process and the resulting fiber structures are summarized in Figure~\ref{fig:3Foldso10}.
From the perspective of the 6D F-theory compactification, the E-string points can be avoided by going to the tensor branch. The blow-up of the ambient variety can be done explicitly by adding the vertices  $ xb_1: (0,1,1,1)$, $ xb_2 :(0,1,1,1)$ to the polytope which gives a new three-fold $Y_{3,c}$ with Hodge numbers
\begin{align}
(h^{1,1},h^{2,1})(Y_{3,c})=(12,40)\,.
\end{align} 
These Hodge numbers are identical to those, of the non-flat model as expected. The resulting fibration has been made fully flat and anomalies can be shown to cancel as expected.
\subsection*{The four-fold compactification}
We follow the same strategy as in Section~\ref{ssec:example1} and fiber the three-fold over another $\mathbb{P}^1$ to obtain the four-fold $Y_{4,a}$. Torically this is done by simply adding a fifth direction in the polytope  \eqref{eq:so10polytope} and by adding the vertices $z_0: (0,0,0,0,1), z_1: (0,0,0,0,-1)$ to fill out the full lattice $\mathbb{Z}^5$. The fibration structure is still inherited from the ambient variety, where the projection is given onto the last three columns resulting in the base $B_3 = (\mathbb{P}^1)^3$.   Again we do not expect to find any three-form cohomology in $Y_{4,a}$ as the fibration is flat in codimension two  and the base $B_3$  too simple to admit $h^{2,1}$ contributions. This is readily checked by computing the Hodge numbers via the Batyrev prescription
\begin{align} 
(h^{1,1}, h^{2,1},h^{3,1})_{\chi}(X_{4,a}) = (11,0,141)_{960}.  
\end{align}  
  We focus again on the structure of the $\mathfrak{so}_{10}$ divisor $x_0=0$  and the $\mathbf{16}_{-\frac14,-\frac12}$ curve over $d_5=0$ which will become the $(4,6,12)$ curve upon deformation. \footnote{From the direct product structure of the base, we simply add to all base classes in $d_i$ in eqn.~\ref{eq:so10classes} a contribution of   $2H_3$.}.
Since   $d_5 \in c_1(B_3)$ we find the $\mathbf{16}_{-\frac14,-\frac12}$ matter curve to be of genus
 \begin{align}
 g_{\mathbf{16}_{(-\frac14,\frac12})}= 1+\frac12 [x_0] \cdot [d_5]\cdot ([x_0]+ [d_5] -  K_b^{-1})=1 \, .
 \end{align}
 For convenience we also give the leading order coefficients of the Weierstrass model in 
  $x_0 =0$ as
 \begin{align}
 \begin{split}
 f&=x_0^3 ( d_1  (2 d_7^2 d_8 - d_6 d_7 d_9 + 2 d_3 d_9^2)+x_0 R_2) +\mathcal{O}(x_0^4) \, , \\
 g&=x_0^4 (d_1^2 d_7^2  d_9^2 + d_1 x_0 R_3)+x_0 d_1)+  \mathcal{O}(x_0^6) \, ,  \\
 \Delta&=x_0^8 ( d_1^4 d_7^4 d_9^4 + x_0 d_1^3 R_i + x_0^2 d_1^2 )  + d_1\mathcal{O}(x_0^{11}) \, ,
 \end{split}
 \end{align}
 with $R_i$ being some residual polynomials. The discriminant can be used to find the perturbative Yukawa and non-perturbative four-point couplings. E.g. there are $(3,5,9)$ points when $d_7=0$ and $d_9=0$ which give rise to the expected trilinear Yukawa couplings of the type
 \begin{align}
\{ x_0=d_5=d_9=0\} \qquad  Y_1: \quad & \mathbf{16}_{-1/4,-1/2 } \cdot \mathbf{16}_{-1/4,-1/2 } \cdot \mathbf{10}_{1/2,1 } \, , \\
\{x_0 = d_5=d_7 =0\} \qquad Y_2:  \quad &  \mathbf{16}_{-1/4,-1/2 } \cdot \mathbf{16}_{-1/4,1/2 } \cdot \mathbf{10}_{1/2,0 }  \, .
 \end{align}
 These couplings are all fully consistent the expected gauge symmetry. In the following we want to focus on the non-perturbative couplings that involve the $\mathbf{16}_{-\frac14,-\frac12}$ matter curve. The only point that involve this curve is localized at
 $x_0=d_1 = d_5= 0$.
   From comparing with the general classes of the matter loci in Table~\ref{tab:SO10Spectrum} and  \eqref{eq:Iimatterloci}  we find the matter curves
 \begin{align*}
\{ x_0=d_1 = d_5= 0\} \quad \ni \mathbf{16}_{-1/4,-1/2 } \, ,  \mathbf{10}_{1/2,0}\, ,  \mathbf{1}_{0,1}\, ,  \mathbf{1}_{-1,-1}  \, ,
 \end{align*}
 to intersect at this point.
 Similar as in the sections before, these loci are to be interpreted as non-perturbative four-point coupling points \cite{Achmed-Zade:2018idx} generated by D1 instantons. 
 Interestingly, this locus allows for two independent four-point couplings that are possible via the involved representations. The two possible coupling are schematically depicted as  
 \begin{align} 
 \mathcal{W} \ni  &\phantom{+} \mathbf{16}_{-1/4,-1/2 } \cdot  \mathbf{16}_{-1/4,-1/2 } \cdot \mathbf{10}_{1/2,0} \cdot \mathbf{1}_{0,1}\, \nonumber \\ & + 
  \mathbf{16}_{-1/4,-1/2 } \cdot  \mathbf{16}_{-1/4,-1/2 } \cdot \mathbf{10}_{-1/2,0} \cdot \mathbf{1}_{-1,-1}  \, . 
 \end{align}
  From the perspective of the smooth geometry, given via \eqref{eq:so10a}, one finds that it is the $f_3=0$ fibral divisor that splits into a non-flat component over $d_1=d_5=0$. This comes as no  surprise as it is going to be the same (ambient divisor) that is pushed to a non-flat surface at codimension two, when performing the conifold transition. 
 \subsection*{Enhancing to non-flat fibers in codimension two}
Now we perform the four-fold analog of the 6D tensor transition which enhances the $\mathbf{16}_{-1/4,-1/2}$ curve over $x_0=d_5=0$ to a (4,6,12) curve via the factorization $d_1 \rightarrow d_1 x_0$, resulting in the four-fold $Y_{4,b}$ 
 This transition again simply merges parts of the $\mathbf{10}$-plet and singlet matter curves into that of  $\mathbf{16}$-over which the fiber attains a non-minimal singularity. In the smooth geometry the non-minimal singularity over that curve is  resolved by the $f_3$ non-flat surface, analogous to the three-fold case. In this regard it makes again sense to interpret the non-flat matter curve as the compactification of 6D E-string theories. 
 
    From the perspective of toric geometry the four-fold $Y_{4,b}$ is simply obtained by adding the same toric vertex $f_1=(0,2,1,0,0)$  to the 5d polytope with CY hypersurface $X_{4,b}$ which admits the Hodge numbers
\begin{align} 
(h^{1,1}, h^{2,1},h^{3,1})_{\chi}(Y_{4,b}) =(12,1,133)_{912} \, .  
\end{align}  
The above data is exactly as expected: First there is the new class that contributes that non-flat surface at $f_3=0$ over the genus-one curve $x_0=d_5=0$ which also contributes in $h^{2,1}(Y_{4,b})$.   
Finally we want to get rid of this non-flat fiber again be performing a base change of   $B_3=(\mathbb{P}^1)^3$.
In the first step we want to deform the curve $x_0=d_5=0$ by performing a complex structure deformation in the $d_5$ followed by a blow-up. For this we remember that $[d_5] \sim c_1(B_3)$. Hence by a complex structure deformation we enforce first the factorization
\begin{align}
d_5 \rightarrow y_0 y_1 z_0  z_1 \hat{d}_5 \, ,
\end{align}
with $[\hat{d}_5]\sim [2 x_0]$ and hence $[x_0] \cdot [\hat{d}_5]\sim 0$. Therefore the only non-trivial curve classes are when $x_0=0$ intersects the other four components $y_i=0$ and $z_i=0$. By performing   the first blow-up in the base, that is inherited from the three-fold $Y_{3,c}$ we are adding the rays  
\begin{align} 
\begin{array}{cc}
 xb_1 : (0,1,  1, -1,  0)\,  , \qquad 
 xb_2 :  (0,1 ,  1,  1,  0)\,  ,
\end{array}  
\end{align}
to the full 5D polytope. Note that the base has become  $B_3=(BL_2 F_0) \times \mathbb{P}^1$ and admits a triangulation with Stanley-Reisner ideal:
\begin{align}
SRI: \{ x_0 x_1, x_0 y_0, x_0 y_1, z_0 z_1, x_1 xb_1, y_1 xb_1, xb_1 xb_2, x_1 xb_2, y_0 xb_2, y_0 y_1 \} \, .
 \end{align} 
 This base change has turned the genus one curve  $x_0=d_5=0$ into two $\mathbb{P}^1$s over $x_0 = z_0=0$ and $x_0 = z_1=0$ that do not intersect and host a non-flat surface each.  The resulting fourfold $Y_{4,c}$ admits the Hodge numbers
\begin{align}
(h^{1,1}, h^{2,1},h^{3,1})_{\chi}(Y_{4,c})= (15,0,129)_{912} \, .
\end{align}
The change in the Hodge numbers is explained as before: First there are two more classes from the two blow-ups which has removed all $h^{2,1}$ contributions. This genus one curve has split into two $\mathbb{P}^1$ with a non-flat fiber over each of them.  Both of these fibers contribute an independent non-flat fiber, which is responsible for the third additional K{\"a}hler parameter. If we want to also get rid of those we need a base that forbids the intersections $x_0=z_i=0$ as well which can be done with yet two more blow-ups. The two additional rays that are to be added to the polytope that do the job are given as
\begin{align} 
 xb_3 :   ( 1,  0,  1, 0,  1) \, , \qquad 
 xb_4 :   ( 1,  0,  1, 0,  -1) \, . 
\end{align}
The resulting four-fold $Y'_{4,c}$ is finally fully flat and admits the very same Hodge numbers as $Y_{4,c}$
\begin{align}
(h^{1,1}, h^{2,1},h^{3,1})_{\chi}(Y'_{4,c})=(h^{1,1},h^{2,1},h^{3,1})_{\chi}(Y_{4,c})=(15,0,129)_{912} \, ,
\end{align} 
We find no change in the Hodge numbers at all, which can be explained by the fact that the non-flat fibers are simply exchanged for the two new base classes.  The blown-up base admits a regular start triangulation with the following SRI 
\begin{align}
\mathcal{SRI}_{B_3}: \{&   \underline{ x_0 y_0, x_0 y_1, x_0 z_0, x_0 z_1},x_0 x_1, , x_1 xb_3, y_0 xb_3, y_1 xb_3, z_1 xb_3,xb_3 xb_4, x_1 xb_1 \nonumber \\ &,  y_1 xb_1, xb_1 xb_2, x_1 xb_4, y_0 xb_4, y_1 xb_4, z_0 xb_4, x_1 xb_2,  y_0 xb_2, y_0 y_1, z_0 z_1 \} \, .
\end{align}
In the above ideal we have underlined the components that forbid the non-flat curve classes when compared to the blow-down. 
We conclude by making the same important observation as in the example from Section~\ref{ssec:example1}: In the transition from $Y_{4,b}$ to $Y'_{4,c}$ all Hodge numbers and in particular $h^{2,1}$ change. However the change is always such that the Euler numbers stay inert. 
This observation is again important when including $G_4$ fluxes and to satisfy the condition in eqn.~\eqref{eq:tadpolechange} which we will comment on in Section~\ref{sec:five}. 
\section{High rank cases: $E_8 \times SU(2) \times SU(3)$ }
\label{sec:four}
In this section we want to consider more complicated examples that admit multiple curves with different superconformal matter curves over each to demonstrate the validity of eqn.~\eqref{eq:HodgeNF4fold}.
  For this we engineer a generalized Tate model, with an $\mathfrak{e}_8$, an $\mathfrak{su}_2$ and an $\mathfrak{su}_3$  divisor via
\begin{align}
p=b_1 Y^2 + b_2 X^3 + a_1 X Y Z + a_2 Z^2 X^2 + a_3 Z^3 Y + a_4 Z^4 X + a_6 Z^6 \, .
 \end{align}
Here, the Arten Tate-algorithm can be used to bring the above model into Weierstrass form  which makes it easier to read off the singularity structure. The Weierstrass coefficients are given as
\begin{align}
\label{eq:tateGeneral}
\begin{split}
f=&	-\frac{1}{48}(a_1^2 - 4 a_2 b_1)^2 + 
 \frac12 b_1 (-a_1 a_3 + 2 a_4 b_1) b_2 \, ,\\ 
 g=& \frac{1}{864} ((a_1^2 - 4 a_2 b_1)^3 + 
   36 b_1 (a_1^2 - 4 a_2 b_1) (a_1 a_3 - 2 a_4 b_1) b_2 + 
   216 b_1^2 (a_3^2 - 4 a_6 b_1) b_2^2)\, .
   \end{split}
\end{align}
The zero-section is given by $Z=0$ in the generalized Tate-model is not a trivial section of $B_3$ anymore. This results in the pre-factors  $b_1$ and $b_2$ to be 
non-trivial divisors of the base whose zero-locus gives the additional $\mathfrak{su}_3$ and $\mathfrak{su}_2$ singularities respectively. The base classes of $b_1$ and $b_2$ parametrize two new line bundle classes  denoted by $\mathcal{Z}_{su3}$ and $\mathcal{Z}_{su2}$. Engineering the additional $\mathfrak{e}_8$ singularity over $\mathcal{Z}_{e8}: z=0$ in Tate form is standard and can be taken from the literature (e.g. \cite{Katz:2011qp}) by factorizing 
\begin{align}
\{ a_1,a_2,a_3,a_4,a_6 \} \rightarrow \{z  a_{1,1},z^2 a_{2,2},z^3  a_{3,3},z^4 a_{4,4},z^5 a_{6,5}  \} \, .
 \end{align} 
The option to allow for $\mathcal{Z}_{su_2}$ and $\mathcal{Z}_{su_3}$ shifts the classes $a_i$  such that one ends up with
\begin{align} 
\label{eq:e8su3su2sec}
\begin{split}
 & [b_1]\sim \mathcal{Z}_{su_3}  \, , \quad 
[b_2]\sim\mathcal{Z}_{su_2}  \, , \qquad   \qquad\qquad
[a_{1,1}] \sim  c_1 - \mathcal{Z}_{e_8}\, ,  \\
&   [a_{2,2}] \sim 2 c_1 - \mathcal{Z}_{su_3} - 2 \mathcal{Z}_{e_8} \, , \, \, \,  \, \,  \quad \qquad\qquad
    [a_{3,3}] \sim 3 c_1 - \mathcal{Z}_{su_2} - \mathcal{Z}_{su_3} - 3 \mathcal{Z}_{e_8}\, ,    \quad  \\
  &  [a_{4,4}]\sim 4 c_1 - \mathcal{Z}_{su_2} - 2 \mathcal{Z}_{su_3} - 4 \mathcal{Z}_{e_8}\, , \qquad   
 [a_{6,5}] \sim 6 c_1 - 2 \mathcal{Z}_{su_2} - 3 \mathcal{Z}_{su_3} - 5 \mathcal{Z}_{e_8} \,  .
 \end{split}
\end{align}  
The model above admits three types of superconformal matter collisions at codimension two which can be read off from eqn.~\eqref{eq:tateGeneral} at
\begin{align}
\label{eq:nonflate8}
\begin{split}
[\text{E-string}]:&\quad z =  a_{6,5} =0 \, , \\ 
[e_8su_2]:&\quad z =  b_{2} =0 \, , \\ 
[e_8su_3]:&\quad z =  b_{1} =0 \, , 
\end{split}
\end{align}
as e.g. discussed in \cite{Apruzzi:2018nre,Apruzzi:2019vpe}. In general  $[e_8 su_n]$ conformal matter gives rise to a $\text{ord}_{\text{van}}(f,g,\Delta)=(4,6,12+n)$ collisions in the Weierstrass model. The tensor branches of those models have been analyzed in detail in \cite{DelZotto:2014hpa}. They are given via a linear chain of $\mathbb{P}^1$'s with self-intersection $(-n)$ given as
\begin{align}
\label{eq:tensore8sun}
\begin{picture}(0,20)
\put(-75,0){\includegraphics[scale=0.55]{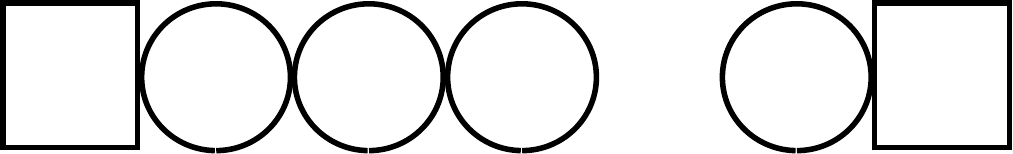}}
\put(-70,10){$\mathfrak{e}_8$}
\put(68,10){$\mathfrak{su}_n$}
\put(-22,26){$\mathfrak{su}_1$}
\put(1,26){$\mathfrak{su}_2$}
\put(40,26){$\mathfrak{su}_{\text{n-1}}$}
\put(-45,9){-1}
\put(-20,9){-2}
\put(5,9){-2}
\put(47,9){-2}
\put(23,10){$\ldots$}
\put(90,10){$\, .$}
\end{picture}  
\end{align} 
The superscript denotes  additional $\mathfrak{su}_k$ gauge algebra factors hat are enforced in the Weierstrass model upon the base change. Note that bi fundamental matter lies in between these nodes.
  These codimension two collisions can be resolved by a non-flat fiber. From the $n$-dimensional 6D tensor branch and upon collecting the contributions of the gauge algebra factors the full 5D coulomb branch dimension is given as
\begin{align}
\label{eq:5De8sun}
\text{dim}\left(\text{Coulomb}_{5D}\left( \left[ \mathfrak{e}_8 \mathfrak{su}_n \right]\right)\right) = 1 +\frac12 n(n-1) \, .
\end{align}  
\begin{table}[t!]
 	\begin{tikzpicture}[scale=1.3] 
 		\node (A) at (0.3,0) [draw,rounded corners,very thick,text width=13cm,align=center] { 
\begin{tabular}{cc}
\begin{tabular}{|c|c|}\hline
 fiber &  ray  \\ \hline
$Z$& (-2, -1, 0, 0, 0) \\
$X$& (1, -1, 0, 0, 0) \\ 
$Y$&(0, 1, 0, 0, 0) \\
$e_1$ & (0, -1, 0, 0, 0) \\ 
$e_2$ & (-1, -1, 0, 0, 0)\\
$e_3$ & (-1, 0, 0, 0, 0) \\ \hline 
\multicolumn{2}{c}{}\\ \hline
$E_8$ -coord & ray \\ \hline
$m_1$ & (-2, -1, 0, 6, 0) \\ 
$l_1$ & (-2, -1, 0, 5, 0) \\
$k_1$ & (-2, -1, 0, 4, 0) \\
$k_2$ & (-1, -1, 0, 4, 0) \\
$h_1$&(-2, -1, 0, 3, 0)\\
$h_2$ &  (-1, 0, 0, 3, 0)\\ 
$g_1$ & (-2, -1, 0, 2, 0)\\
$g_2$ &  (0, -1, 0, 2, 0) \\ 
$f_1$ & (-2, -1, 0, 1, 0) \\ \hline 
\end{tabular}  
&
\begin{tabular}{|c|c|}\hline
 surface  &  ray  \\ \hline 
$h_3$ & (-1, -1, 0, 3, 0)\\
$f_2$ &   (-1, -1, 0, 1, 0) \\
$f_3$ &  (0, -1, 0, 1, 0) \\ 
$g_4$& (-1, -1, 0, 2, 0) \\ \hline 
$f_4$ & (-1, 0, 0, 1, 0) \\ 
$g_3$ & (-1, 0, 0, 2, 0) \\ \hline
$f_5$ &  (0, 0, 0, 1, 0)  \\ \hline  
\multicolumn{2}{c}{}\\ \hline
Base & ray \\ \hline
  $x_0$ &  ($n_{su_2}$+$n_{su_3}$-2, $n_{su_3}$-1, -1, -1, -1)  \\ $x_1$ &  (-2, -1, 1, 0, 0)  \\ $x_3 $&  (-2, -1, 0, 0, 1)  \\ \hline 
\end{tabular} 
\end{tabular}  
};
 \node (B) at (0.3,-8)  { 	   
 	    $
 \begin{array}{|c|c|c|c|c c|}\hline
 h^{1,1} & h^{2,1} & h^{3,1} & \chi & n_{su_2} & n_{su_3} \\ \hline 
  11& 153& 1982 & 11088& 
 0&  0\\  
   17& 105& 1300 & 7320&  
0&  1\\ 
   17& 66& 801 & 4560&  
0&  2\\ 
   17& 40& 459 & 2664&  
 0&  3\\ 
   17& 27& 250 & 1488& 
0&  4\\ 
  17& 27& 150 & 888&  
0&  5\\ 
   17& 40& 135 & 720&  
 0&  6\\ 
   14& 120& 1522 & 8544&  
1&  0\\ 
   20& 78& 960 & 5460&  
1&  1\\ 
   20& 45& 563 & 3276&  
1&  2\\ 
   20& 25& 305 & 1848&  
1&  3\\ 
   20& 18& 162 & 1032&  
1&  4\\ 
   20& 24& 110 & 684&  
1&  5\\ 
  14& 91& 1147 & 6468&  
2&  0\\ 
   20& 55& 693 & 3996&  
2&  1\\ 
   20& 28& 386 & 2316&  
2&  2\\ 
   20& 14& 200 & 1284&  
2&  3\\ 
   20& 13& 111 & 756&  
2&  4\\ 
   19& 24& 95 & 588&  
2&  5\\ 
   14& 68& 850 & 4824&  
3&  0\\  \hline
\end{array} \qquad 
 \begin{array}{|c|c|c|c|c c|}\hline
 h^{1,1} & h^{2,1} & h^{3,1} & \chi & n_{su_2} & n_{su_3} \\ \hline
  20& 38& 492 & 2892&  
3&  1\\ 
  20& 17& 263 & 1644&  
3&  2\\ 
   20& 9& 137 & 936&  
 3&  3\\    20& 14& 90 & 624&  
 3&  4\\ 
  14& 51& 625 & 3576&  
4&  0\\ 
   20& 27& 351 & 2112&  
4&  1\\ 
   20& 12& 188 & 1224&  
  4&  2\\ 
   20& 10& 110 & 768&  
  4&  3\\ 
  14& 40& 466 & 2688&  
  5&  0\\ 
   20& 22& 264 & 1620&  
 5&  1\\ 
   20& 13& 155 & 1020&  
  5&  2\\ 
   19& 16& 113 & 744&  
  5&  3\\ 
   14& 35& 367 & 2124&  
  6&  0\\ 
  20& 23& 225 & 1380&  
 6&  1\\ 
   20& 20& 158 & 996&  
  6&  2\\ 
   14& 36& 322 & 1848&  
  7&  0\\ 
  20& 30& 228 & 1356&  
 7&  1\\ 
   14& 43& 325 & 1824&  
  8&  0\\ 
   19& 42& 267 & 1512&  
 8&  1\\ 
   14& 56& 370 & 2016&  
 9&  0\\ \hline
 \end{array}  $}; \end{tikzpicture}
 \caption{{\it \label{tab:polye8su3su2}Summary of the rays of the polytope $\Delta (n_{su_3},n_{su_2})$ that leads to  smooth elliptic four-fold $X_4 (n_{su_3},n_{su_2})$ with $\mathfrak{e}_8$, $\mathfrak{su}_3$ and $\mathfrak{su}_2$ fibers over a base $\mathbb{P}^3$. The Hodge numbers of all consistent inequivalent four-folds are summarized below.}}  
\end{table}  
As argued in Section~\ref{ssec:4612}  the non-flat fiber resolution must include the same amount of surface contributions with their own K{\"a}hler classes. In order to compute the contribution to $h^{2,1}$ as well, requires the computation of the genus of those curves in $B_3$, given as 
\begin{align}
\begin{split}
g_{\text{E-string}}=& 1 + \frac12 (6 c_1 - 2 \mathcal{Z}_{su_2} - 3 \mathcal{Z}_{su_3} - 5 \mathcal{Z}_{e_8})\cdot  (5 c_1 - 2 \mathcal{Z}_{su_2} - 3 \mathcal{Z}_{su_3} - 4 \mathcal{Z}_{e_8}) \cdot \mathcal{Z}_{e_8} \, , \\ 
g_{e_8su_2}=&1 + \frac12 \mathcal{Z}_{su_2} \cdot \mathcal{Z}_{e_8}\cdot ( \mathcal{Z}_{su_2} + \mathcal{Z}_{e_8}-c_1) \, , \\
g_{e_8su_3}=&1 + \frac12 \mathcal{Z}_{su_3}\cdot  \mathcal{Z}_{e_8}\cdot  (  \mathcal{Z}_{su_3} + \mathcal{Z}_{e_8}-c_1)\, .
\end{split}
\end{align}
Putting all pieces together, the contributions
  to $h^{2,1}(X_4)$ via eqn.~\eqref{eq:HodgeNF4fold} and eqn.~\eqref{eq:nonflath11} and using the knowledge of the tensor branches in eqn.~\eqref{eq:5De8sun} is given as
\begin{align}
h^{2,1}_{\text{non-flat}}(X_4)=g_{\text{E-string}} + 2 g_{[e_8 su_2]}+4 g_{[e_8 su_3]} \, .
\end{align}
This can be explicitly checked by  constructing a family of four-folds that exhibit a simple toric description. The simplest base to take is $B_3=\mathbb{P}^3$. In terms of the base hyperplane class $H_1$ we can fix all line bundle choices as
\begin{align}
\mathcal{Z}_{e8} \sim H \, , \quad \mathcal{Z}_{su_3} \sim n_{su_3} H \, ,  \quad \mathcal{Z}_{su_2} \sim n_{su_2} H \, . 
\end{align}
Here the $n_{su_n}$ are positive integers bounded  by effectiveness of all  $b_i, a_{i,k}$ in \eqref{eq:e8su3su2sec}. For those constructions, the base is again too simple to contribute $h^{2,1}$.  It is also important to note that this particular geometry does not come from a direct compactification of an already present 6D theory.  Hodge numbers of the family of four-folds $X_4(n_{su_2},n_{su_3})$ are given as
\begin{align}
\label{eq:hodgee8} 
h^{1,1}(X_4(n_{su_2},n_{su_3})) =& 10+ (1-\delta_{(19-2n_{su_2}-3n_{su_3},0)})+ 3(1- \delta_{(n_{su_2},0)}) + 6(1- \delta_{(n_{su_3},0)})  \nonumber \, , \\
h^{2,1}(X_4(n_{su_2},n_{su_3})) =&  g_{\text{E-string}} (1-\delta_{(19-2n_{su_2}-3n_{su_3},0)})  \nonumber \\   +& 2g_{e_8su_2} (1- \delta_{(n_{su_2},0)}) + 4g_{e_8 su_3}(1- \delta_{(n_{su_3},0)})   \, . 
\end{align}
This is double checked via the Batyrev construction. The 5D polytope $\Delta (n_{su_3},n_{su_2})$ that realizes the two-parameter family of four-folds  $X_4 (n_{su_3},n_{su_2})$ is summarized in Table~\ref{tab:polye8su3su2}. In the same table also Hodge and Euler numbers are computed for all reflexive polytopes consistent with the expectation from eqn.~\eqref{eq:hodgee8}.
The fully resolved hypersurface is given as 
\begin{align}
\begin{split}
p=\phantom{+}&b_1 e_3 f_4 f_5 g_3 h_2 Y^2 +b_2 e_1^2 e_2 f_2 f_3^2 f_5 g_2^2 g_4 h_3 k_2 X^3  \\+ & 
 a_{1,1} e_1 e_2 e_3 f_1 f_2 f_3 f_4 f_5 g_1 g_2 g_3 g_4 h_1 h_2 h_3 k_1 k_2 l_1 m_1 X Y Z   \\ +&  
 a_{2,2} e_1^2 e_2^2 e_3 f_1^2 f_2^2 f_3^2 f_4 f_5 g_1^2 g_2^2 g_3 g_4^2 h_1^2 h_2 h_3^2 \
k_1^2 k_2^2 l_1^2 m_1^2 X^2 Z^2   \\ +& 
 a_{3,3} e_1 e_2^2 e_3^2 f_1^3 f_2^2 f_3 f_4^2 f_5 g_1^3 g_2 g_3^2 g_4^2 h_1^3 h_2^2 \
h_3^2 k_1^3 k_2^2 l_1^3 m_1^3 Y Z^3   \\ +& 
 a_{4,4} e_1^2 e_2^3 e_3^2 f_1^4 f_2^3 f_3^2 f_4^2 f_5 g_1^4 g_2^2 g_3^2 g_4^3 h_1^4 \
h_2^2 h_3^3 k_1^4 k_2^3 l_1^4 m_1^4 X Z^4   \\ +& 
 a_{6,5} e_1^2 e_2^4 e_3^3 f_1^5 f_2^3 f_3 f_4^2 g_1^4 g_3 g_4^2 h_1^3 h_3 k_1^2 l_1 Z^6 \, ,
 \end{split}
\end{align} 
and has been analyzed for various (fiber) triangulations e.g. in \cite{Apruzzi:2018nre} with $f_1$ being the affine component of $\mathfrak{e}_8$. The split into the various non-flat loci at \eqref{eq:nonflate8} can be readily verified at the loci given in eqn.~\eqref{eq:nonflate8}.  
\section{Conclusion and Outlook}
\label{sec:five}
This note has considered a systematic analysis of smooth elliptic four-folds that exhibit non-flat fibers. First we have shown how non-flat fibers in codimension two contribute to the Hodge numbers and in particular to the three-form cohomology.
Via the M-theory duality such three-forms lead to additional chiral singlets. In F/M-theory these non-flat configurations are to be interpreted as compactifications of 6D/5D superconformal matter theories on a Riemann-surface. This allowed us to identify their contributions to the Hodge numbers in terms of the 6D/5D tensor(coulomb) branch dimensions and the genus of the Riemann-surface. The validity of this proposal is checked for several examples which include 6D SCFTs of various ranks.\\
Furthermore we have investigated conifold transitions among four-folds that remove those non-flat fibers and hence these specific chiral singlet fields. The first branch of these transitions is analogous to a 6D tensor branch  and corresponds to a birational base change. This transition changes all Hodge numbers but most notably does not change the Euler number of the compact model. Note that this work has focused primarily on the geometric aspects and not the inclusion of $G_4$ flux. The fact that the Euler number does not change in such transitions simplifies the matching of 4D SUSY vacua substantially \cite{Intriligator:2012ue,Jockers:2016bwi} due to the vanishing of the right hand side of 
\begin{align}
\label{eq:tadpolechange}
\Delta \left(n_{D3} \right)  + \frac12 \Delta \left(\int G_4 \wedge G_4 \right)= \Delta \left( \frac{\chi}{24} \right) \, .
\end{align} 
Euler number preserving transitions therefore do not require a to change the $G_4$ flux (norm) or number of $D3$ branes during the transition.
  The second type of transitions we have considered is analogous to a (partial) 6D Higgs branch that keeps the total gauge group but moves the non-flat fiber from curves down to points of the base $B_3$. These points lead to non-perturbative four-point matter couplings, mediated by D1 instantons and do change the Euler number. The existence of these non-perturbative interaction points is enforced geometrically and can be interpreted as a remnant of the 6D E-string theory. In fact we have argued that the matter representations involved in such couplings can be deduced due to the anomalies of the 6D E-string transition.  

This work serves as a first step towards the investigation of 4D/3D theories obtained from F/M-theory on non-flat elliptic four-folds. From here on there are several directions to go in the future. First it is important to fully include $G_4$ fluxes. As argued, the right starting point might be the phase without those fibers compute a consistent configuration and perform the conifold to the non-flat configuration.
   Moreover in order to  fully understand  the 4D theory and its possible non-perturbative effects, a better understanding of the 4D effective action is desirable. As analyzed in \cite{Apruzzi:2018oge}, Euclidean D3 instantons can lead to a mixing of the complex structure and K{\"a}hler moduli. This work shows that three-forms are a potential third contribution to those moduli sectors that are characteristic to non-flat resolution. Understanding all those effects is important to clarify how quantum corrections might obstruct 4D SCFT points in the IR.  As the four-fold is fully smooth those contributions are best analyzed in the 3D M-theory.
There are also further generalizations possible from pure geometric point of view.
 These include the addition of monodromies that act on the superconformal matter curves, analogous to to split fibers in elliptic three-folds. Such effect would naturally incorporate the folding action familiar from twisted compactifications \cite{Bhardwaj:2019fzv} into the four-dimensional picture. 
Finally it would also be very interesting to explicitly construct the heterotic duals of those theories and match the contributions of $h^{2,1}_{\text{non-flat}}(X_4)$ to that of the NS5 branes \cite{Braun:2018ovc}.    
\section*{Acknowledgements}
The author thanks Markus Dierigl,  Mohsen Karkheiran, Magdalena Larfors, Fabian Ruehle and Thorsten Schimannek for discussions. P.O. thanks in particular Markus Dierigl and Thorsten Schimannek  for reading an earlier version of the draft and their useful comments.
The author would also like to acknowledge the hospitality of the SCGP during the program ``Geometry and Physics of Hitchin Systems'' where initial parts of this work.
The work of P.K.O.\ is supported by a grant of the Carl Trygger Foundation for Scientific Research.
\appendix
\section{Toric resolution of the $(E_6 \times U(1))/\mathbb{Z}_3$ model}
\label{app:E6toric}
In this appendix we look at some geometric details that were left out in the main section. These include in particular the fully resolved three-and fourfolds and their intersections. 
\subsection{The resolved Tate model} 
Starting from the tuning of the Tate model in Section~\ref{ssec:example1} we need to resolve the $I_2$ via the exceptional divisor $e_1=0$ the $\mathfrak{e}_6$ with the set $\{f_0, f_1, f_2, g_1 , g_2, g_3, h_1\}$ that can be engineered as blow-ups of the $\mathbb{P}^{1,2,3}$ fiber ambient space. The fully resolved U(1) restricted Tate model is given by
\begin{align}
p=&  e_1 f_1 g_3 Y^2   + 
 e_1^2 f_1 f_2^2 g_2 X^3 +a_{1,1} e_1 f_0 f_1 f_2 g_1 g_2 g_3 h_1 X Y Z \nonumber \\ & + 
 a_{2,2} e_1 f_0^2 f_1 f_2^2 g_1^2 g_2^2 g_3 h_1^2 X^2 Z^2 + a_{3,2} f_0^2 g_1 Y Z^3 + 
 a_{4,3} f_0^3 f_2 g_1^2 g_2 h_1 X Z^4 
\end{align}
There are two sections given by $Z=0$ that intersects the affine node $f_0$ and $e_1=0$ that intersects $f_2$ in a phase that employs the SRI:
\begin{align}
SRI: \{&  Z e_1, Z f_1, Z f_2, Z g_1, Z g_2, Z g_3, Z h_1, Y X, Y f_2, Y g_2, Y h_1,   e_1 f_0, f_0 f_1, f_0 g_2, f_0 g_3, \nonumber \\ &f_0 h_1, X f_1, X g_1, X g_2, X g_3, X h_1, e_1 g_1, e_1 g_2, e_1 g_3, e_1 h_1, f_1 g_1, f_1 h_1, f_2 g_3, f_2 h_1 \}
  \end{align}
Graphically we give the intersection of the $\mathfrak{e}_6^{(1)}$ fiber as 
  \begin{align*}
  \begin{picture}(0,55)
  \put(-10,10){\includegraphics[scale=2.5]{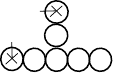}} 
  \put(39,50){$[f_0]$}
    \put(39,35){$[g_1]$}
        \put(-12,0){$[f_2]$}
         \put(-7,35){$e_1$}
        \put(10,50){$Z$}
        \put(5,0){$[g_3]$}
        \put(22,0){$[h_1]$}
         \put(38,0){$[g_2]$}
        \put(55,0){$[f_1]$}
         \put(80,20){$.$}
  \end{picture} 
  \end{align*}
  The intersections allow us to compute the $U(1)$ charges by using the smooth geometry in M-theory and lift it back  to F-theory. For this we employ the Shioda map, that is the divisor that gives the $U(1)$ generator in the M-theory expansion of the $C_3$ form. For simplicity we focus here on the fibral part, which is given in the difference of the two sections. 
  By then demanding orthogonality with respect to all other fibral divisors leaves us with the following linear combination
\begin{align}
\label{eq:shiodae6}
\sigma(s_1) = [e_1]-[Z]+ \frac13(4 f_1 +5 g_3 +6  h_1 +4 g_2 +2 f_2 +3 g_1) \,.
\end{align}
The above concrete form allows us to compute the explicit charges of matter multiplets by intersecting it with the reducible fibral curves at codimension two. E.g. first there are the  $\mathbf{1}_1$ over $a_{4,3}=a_{3,2}=0$. Here one can see, that the coordinate $e_1=0$ becomes reducible. The fiber topology is that of an $I_2$ curve as given in Figure~\ref{fig:4FoldE61}. Since $e_1=0$ is a $-1$ curve on the fiber ambient space which coincides with the $U(1)$ charge of the singlets using eqn.\eqref{eq:shiodae6}. Then there are the
   $\mathbf{27}_{\frac13}$-plets over $z=a_{3,2}=0$. This can be seen by noting, that the fiber here becomes of $\mathfrak{e}_7$ type. From the reducible fiber components one can deduce the $\mathfrak{e}_6$ charges using \eqref{eq:shiodae6}. For four-folds there is an additional enhancement locus that is of interest. A codimension three by first setting $a_{3,2}=a_{4,3}=0$, where the hypersurface becomes
\begin{align}
p=e_1 f_1 ( g_3 Y^2  + 
   e_1 f_2^2 g_2 X^3 +a_{1,1} f_0 f_2 g_1 g_2 g_3 h_1 X Y Z + 
   a_{2,2} f_0^2 f_2^2 g_1^2 g_2^2 g_3 h_1^2 X^2 Z^2   ) \, ,
\end{align}
This is also the $\mathbf{1}_1$ singlet locus, but note that the $f_1$ $\mathfrak{e}_6$ component also factors out. As $f_1=0$ restricts onto the $z=0$ locus in the base, setting it to zero is a codimension three locus. Moreover, coordinates $\{ f_2, g_2,g_3 , Y, e_1 \}$  are fully unrestricted and they parametrize a two-dimensional surface. In Figure~\ref{fig:4FoldE61} we have computed all intersections of the fibral components over that locus. In the red box we have computed the intersections of the five components that sit inside the non-flat surface $f_1=0$. This is unlike the shape a regular\footnote{In codimension three it appears often, that the fiber obtains a bouquet shape form, where certain nodes are deleted. This is however } Dynkin diagram shaped fiber should have but it is characteristic for a non-flat fiber.  
Secondly we the enhanced model with a non-flat fiber at codimension two. This is done, by further factorizing one power of $z$ out of $a_{4,3}$ as $z b_{4,4}$. This leads to the codimension two non-flat curve over $z=a_{3,2}=0$. Note that we use the toric resolution of the fiber as given via the top construction that we are using to construct full three-and four-folds in the next section. We have added the new fibral coordinate $f_3$ which removes $f_1$ as one of the $\mathfrak{e}_6$ fibral divisors. Notably, it is still present in the ambient space and will be introduced as a non-flat fiber momentarily. The new resolved fibration is given as
\begin{align}
p= &\phantom{+}e_1 f_1 f_3^2 g_3 Y^2  +  
 e_1^2 f_1 f_2^2 g_2 X^3  + a_{1,1} e_1 f_0 f_1 f_2 f_3 g_1 g_2 g_3 h_1 X Y Z   \nonumber \\ &+
 a_{2,2} e_1 f_0^2 f_1 f_2^2 g_1^2 g_2^2 g_3 h_1^2 X^2 Z^2 +_{3,2} f_0^2 g_1 Y Z^3 + 
 b_{4,4} f_0^4 f_1 f_2^2 g_1^4 g_2^3 g_3^2 h_1^4 X Z^4  \, .
\end{align}
Intersections can be computed via a triangulation of the toric ambient space, that leads to the SRI:
\begin{align}
\mathcal{SRI}: \{&  Z e_1, Z f_1, Z f_2, Z g_1, Z g_2, Z g_3, Z h_1, Z f_3, Y X, Y f_1, Y f_2, 
Y g_2, Y h_1, e_1 f_0, f_0 f_1 \nonumber \\ &, f_0 g_2, f_0 g_3, f_0 h_1, f_0 f_3, X g_1, X g_2, X g_3, 
X h_1, X f_3, e_1 f_2, e_1 g_1, e_1 g_2, e_1 g_3, \nonumber \\ & e_1 h_1, f_2 f_3, g_1 f_3, g_2 f_3, 
h_1 f_3, f_1 g_1, f_1 h_1, f_2 g_3, f_2 h_1 ,g_1 g_2 g_3 \}
\end{align}
All intersections of the $\mathfrak{e}_6$ fiber components as well as the $\mathbf{1}_1$ singlets can be computed as usual and their structure has not changed at all. Notably, over $a_{3,2}=0$ the fiber becomes
\begin{align}
P=f_1 (&e_1^2 f_2^2 g_2 X^3 + e_1 f_3^2 g_3 Y^2  + 
   a_{1,1} e_1 f_0 f_2 f_3 g_1 g_2 g_3 h_1 X Y Z   \nonumber  \\ &\quad +
   a_{2,2} e_1 f_0^2 f_2^2 g_1^2 g_2^2 g_3 h_1^2 X^2 Z^2 + 
   b_{4,4}f_0^4 f_2^2 g_1^4 g_2^3 g_3^2 h_1^4 X Z^4)\, .
\end{align}
We have exactly the same behavior as before, that is $f_1=0$ is a non-flat fiber component but this time it happens
over codimension two. Note also that it is exactly the same ambient space coordinate the appears here. Using the SRI above, one can also compute the intersections of the fibral components as given in Figure~\ref{fig:4FoldE62}. There we observe again a similar intersection picture as in codimension three, that is a loop of components that sit inside the non-flat surface.
\subsection{Explicit Toric three-and four-fold}
We give the explicit toric polytope for the three-folds and four-folds that are considered in Section~\ref{sec:three}.
The toric rays of the polytope are given in the tables below. The threefold $X_3$ is constructed from a regular fine star triangulation of the toric fan associated to the Batyrev polytope $\Delta$. These polytopes are a combination of various rays as given below. The projection onto the bases $B$ is inherited from the toric ambient space and we fix it to be onto the last two coordinates. We then give the various polytope ingredients below.   The U(1) restricted Tate-model is a combination of the  $\mathfrak{e}_6$ {\it top } and completed with the choice of a base. The $\mathfrak{e}_6$ resolution divisor becomes non-flat upon adding the ray $f_1$ and the non-flat fibers can be avoided by adding the two $-1$ curves that correspond to the toric rays $b_1$ and $b_2$ to the base. Note that these are next to the $-4$ curve on which the $\mathfrak{e}_6$ is localized and thus reduces its self-intersection by one each time. 
  \begin{center} 
   \begin{tabular}{cccc}
  \begin{tabular}{|c|c|}
   \multicolumn{2}{c}{F: $U(1)$-Tate fiber }\\ \hline
  X & (1,-1,0,0)\\
  Y & (0,1,0,0) \\
  Z & (-2,-1,0,0) \\ \hline
  $e_1$ & (1,0,0,0) \\ \hline 
  \end{tabular}
  &
    \begin{tabular}{|c|c|} 
    \multicolumn{2}{c}{E1: $\mathfrak{e}_6$ Fiber}\\ \hline
 $ f_0$ & (-2,-1,0,-1)\\
 $ f_1  $& (0,0,0,-1) \\
 $ f_2 $ & (0,-1,0,-1) \\ 
 $ g_1 $ & (-2,-1,0,-2) \\ 
 $ g_2 $ & (-1,-1,0,-2) \\ 
 $ g_3 $ & (-1,0,0,-2) \\ 
 $ g_3 $ & (-2,-1,0,-3) \\ \hline 
  \end{tabular} 
    &
        \begin{tabular}{|c|c|} 
    \multicolumn{2}{c}{E2: non-flat-$\mathfrak{e}_6$ Fiber}\\ \hline
 $ f_0$ & (-2,-1,0,-1)\\
 $\mathbf{ f_1 }$& (0,0,0,-1) \\
 $ f_2 $ & (0,-1,0,-1) \\ 
 $ g_1 $ & (-2,-1,0,-2) \\ 
 $ g_2 $ & (-1,-1,0,-2) \\ 
 $ g_3 $ & (-1,0,0,-2) \\ 
 $ g_3 $ & (-2,-1,0,-3) \\  
 $ f_4 $ & (0,1,0,-1) \\ \hline 
  \end{tabular} 
    &
    \begin{tabular}{|c|c|} 
    \multicolumn{2}{c}{B1: $\mathbb{F}_4$ }\\ \hline
 $ z_1$ & (-2,-1,0,1)\\
 $ x_0 $ & (-2,-1,1,0) \\
  $x_1$ & (-2,-1,-1,-4) \\ \hline  
   \multicolumn{2}{c}{ } \\ 
    \multicolumn{2}{c}{B2: BL$_2 \mathbb{F}_4$}\\ \hline
 $ z_1$ & (-2,-1,0,1)\\
 $ x_0 $ & (-2,-1,1,0) \\
  $x_1$ & (-2,-1,-1,-4) \\ \hline 
  $b_1$ & (-2,-1,1,-1) \\  
  $b_2$ & (-2,-1,-1,-5) \\ \hline 
  \end{tabular}  
  
  \end{tabular} 
   \end{center}
   The constituents above are constructed to toric polytopes from which we can compute the Hodge numbers via the Batyrev construction. The results in addition to flatness are given below:
   \begin{align}
   \begin{array}{|c|c|c|c|c|}\hline
   \Delta  & CY & h^{1,1} & h^{2,1} & \text{Flat} \\ \hline 
 $  (F,E1,B1) $& X_3 & 10 & 162 & \checkmark \\ \hline
  $ (F,E2,B1)$ & \hat{X}_3 & 12 & 160 & X \\ \hline
 $    (F,E2,B2)$ & \widehat{X}_3 & 12 & 160 & \checkmark \\ \hline
  $   (F,E1,B1)$ & \widehat{X}_3 & 12 & 160 & \checkmark \\ \hline
 \end{array}
   \end{align} 
 Similarly, the four-folds are constructed by enlarging the toric ambient space and adding a $\mathbb{P}^1$ direction. These are given via two new toric rays $y_0$ and $y_1$ in the following. Using the two base blow-ups $b_1, b_2$ from the thee-folds degenerates the curve with the non-flat fiber in codimension two from a genus five curve to six genus zero curves in the base. These transitions change the Hodge numbers but not the Euler number of the fourfold as can be seen directly. To fully resolve those six non-flat $\mathbb{P}^1$'s one needs to introduce the six blow-ups $b_{3\ldots 8}$. These blow-ups destroy the direct product structure of the base but make the fibration fully flat. Note that these last six blow-ups keep all Hodge numbers invariant in full analogy to its 6D cousins. The various polytope ingredients are given below:
  \begin{center}
 { \small
   \begin{tabular}{cccc  }
  \begin{tabular}{|c|c|}
   \multicolumn{2}{c}{F: $U(1)$-Tate fiber }\\ \hline
  X & (1,-1,0,0,0)\\
  Y & (0,1,0,0,0) \\
  Z & (-2,-1,0,0,0) \\ \hline
  $e_1$ & (1,0,0,0,0) \\ \hline 
  \end{tabular}
  &
    \begin{tabular}{|c|c|} 
    \multicolumn{2}{c}{E1: $\mathfrak{e}_6$ Fiber}\\ \hline
 $ f_0$ & (-2,-1,0,-1,0)\\
 $  f_1  $& (0,0,0,-1,0) \\
 $ f_2 $ & (0,-1,0,-1,0) \\ 
 $ g_1 $ & (-2,-1,0,-2,0) \\ 
 $ g_2 $ & (-1,-1,0,-2,0) \\ 
 $ g_3 $ & (-1,0,0,-2,0) \\ 
 $ g_3 $ & (-2,-1,0,-3,0) \\ \hline    \multicolumn{2}{c}{ } \\ 
    \multicolumn{2}{c}{E2: $\mathfrak{e}_6$ Fiber}\\ \hline
 $ f_0$ & (-2,-1,0,-1,0)\\
 $\mathbf{ f_1 }$& (0,0,0,-1,0) \\
 $ f_2 $ & (0,-1,0,-1,0) \\ 
 $ g_1 $ & (-2,-1,0,-2,0) \\ 
 $ g_2 $ & (-1,-1,0,-2,0) \\ 
 $ g_3 $ & (-1,0,0,-2,0) \\ 
 $ g_3 $ & (-2,-1,0,-3,0) \\ \hline
 $ f_4 $ & (0,1,0,-1,0) \\ \hline 
  \end{tabular} 
    &
    \begin{tabular}{|c|c|} 
    \multicolumn{2}{c}{B1: $\mathbb{F}_4 \times \mathbb{P}^1 $ base }\\ \hline
 $ z_1$ & (-2,-1,0,1,0)\\
 $ x_0 $ & (-2,-1,1,0,0) \\
  $x_1$ & (-2,-1,-1,-4,0) \\ \hline  
  $y_0$ & (-2,-1,0,0,1) \\   
  $y_1$ & (-2,-1,0,0,-1) \\ \hline   
      \multicolumn{2}{c}{ }\\  
    \multicolumn{2}{c}{B2: $(\text{BL}_2 \mathbb{F}_4) \times \mathbb{P}^1)$}\\ \hline
 $ z_1$ & (-2,-1,0,1,0)\\
 $ x_0 $ & (-2,-1,1,0,0) \\
  $x_1$ & (-2,-1,-1,-4,0) \\ \hline 
  $b_1$ & (-2,-1,1,-1,0) \\  
  $b_2$ & (-2,-1,-1,-5,0) \\ \hline 
  $y_0$ & (-2,-1,0,0,1) \\  
  $y_1$ & (-2,-1,0,0,-1) \\ \hline 
  \end{tabular}  
    & 
    \begin{tabular}{|c|c|} 
    \multicolumn{2}{c}{B3: Bl$_6(\text{BL}_2 \mathbb{F}_4) \times \mathbb{P}^1)$}\\ \hline
 $ z_1$ & (-2,-1,0,1,0)\\
 $ x_0 $ & (-2,-1,1,0,0) \\
  $x_1$ & (-2,-1,-1,-4,0) \\ \hline 
  $b_1$ & (-2,-1,1,-1,0) \\  
  $b_2$ & (-2,-1,-1,-5,0) \\ \hline 
  $y_0$ & (-2,-1,0,0,1) \\  
  $y_1$ & (-2,-1,0,0,-1) \\ \hline  
  $ b_3 $& (-2,-1,0,-1,1) \\
 $ b_4$ & (-2,-1,0,-1,-1) \\ 
 $ b_5 $ & (-2,-1,0,-2,1) \\ 
 $ b_6 $ & (-2,-1,0,-2,-1) \\ 
 $ b_7 $ & (-2,-1,0,-3,-1) \\ 
 $ b_8 $ & (-2,-1,0,-3,-1) \\ \hline 
  \end{tabular}  
  
  \end{tabular}
  }
   \end{center}
From the data above, the polytope $\Delta$ can be composed that allows an easy computation of Hodge numbers via the Batyrev formula as:
   \begin{align}
   \begin{array}{|c|c|c|c|c|c|c|}\hline
   \Delta  & CY & h^{1,1} & h^{2,1}& h^{3,1}& \chi & \text{Flat} \\ \hline 
 $  (F,E1,B1) $& X_4 & 11 & 0  & 1447 &8769 & \checkmark \\ \hline
  $ (F,E2,B1)$ & \hat{X}_4 & 12 &6& 1281 & 7776& X \\ \hline
 $    (F,E2,B2)$ & \widehat{X}_4 & 19 &0 &  1269& 7776 & X \\ \hline
  $   (F,E2,B3)$ & \widehat{X}^\prime_4 & 19 & 0& 1269 &7779 & \checkmark \\ \hline 
  $   (F,E1,B3)$ & \widehat{X}^\prime_4 & 19 & 0& 1269 &7779 & \checkmark \\ \hline
 \end{array}
   \end{align} 
   
\section{Review: Hodge numbers from polytopes}
\label{app:Batyrev}
The main framework, that we use in order to construct elliptic three-and fourfolds are going to be hypersurfaces in toric varieties. The toric ambient space is encoded by a regular fine start triangulation of the polytope $\Delta$. In particular for fourfolds, the Hodge numbers can be easily computed the polytope \cite{Klemm:1996ts} 
\begin{align}
\label{eq:Hodge4folds}
\begin{split}
h^{1,1}(\Delta) =& l(\Delta) - 6 - \sum_{\text{dim}(\theta)=4 } l^* (\theta)+\sum_{\text{codim}(\theta)=2 } l^*(\theta_i)l^*(\theta_i^*) \, , \\
h^{2,1}(\Delta) =& \sum_{\text{codim}(\theta)=3 } l^*(\theta_i)l^*(\theta_i^*) \, , \\ 
h^{3,1}(\Delta) =& l(\Delta^*) - 6 - \sum_{\text{dim}(\theta*)=4 } l^* (\theta^*)+\sum_{\text{codim}(\theta^*)=2 } l^*(\theta_i^*)l^*(\theta_i) \, ,  
\end{split}
 \end{align} 
with $\theta$ being faces of $\Delta$ and $\delta^*$ of its polar $\Delta^*$. $l$ denotes points and $l^*$ interior ones of the respective facets. The sum in the last terms goes over pairs of dual facets $\theta$.

\end{document}